\documentclass[%
 reprint,
nofootinbib,
 amsmath,amssymb,
 aps,
]{revtex4-2}

\usepackage{gensymb}
\usepackage{amsfonts}
\usepackage{tikz}
\usetikzlibrary{patterns, patterns.meta}
\usepackage{wrapfig}
\usepackage{appendix}
\usepackage{standalone}
\usepackage{graphicx}% Include figure files
\usepackage{dcolumn}% Align table columns on decimal point
\usepackage{mwe}
\usepackage{adjustbox}
\usepackage{subcaption}
\usepackage{bm}% bold math
\usepackage[colorlinks=true, allcolors=blue]{hyperref}

\begin{document}

\preprint{APS/123-QED}

\title{The Crooks relationship in single molecule pulling simulations of coupled harmonic oscillators}

\author{Julián David Jiménez-Paz}
\email{jujimenezp@unal.edu.co}

\author{José Daniel Muñoz-Castaño}%
\email{jdmunozc@unal.edu.co}
\affiliation{%
 Simulations of Physical Systems Group, CoE-SciCo Excellence Center, Department of Physics, Universidad Nacional de Colombia,
 Carrera 45 No. 26-85, Edificio Uriel Gutiérrez, Bogotá D.C., Colombia.
}%

\date{\today}

\begin{abstract}
In this work, we propose two models of coupled harmonic oscillators under Brownian motion to computationally study the applications of fluctuation theorems. This paper also illustrates how to analytically calculate free energy differences for these systems.
The computational results clearly show that Crooks relation is able to predict free energy differences between initial and final canonical ensembles with around $1\%$ accuracy by using probability distributions of cumulative work done during nonequilibrium protocols carried out with velocities up to three orders of magnitude larger than the quasi-stationary evolution velocity. The curves of instantaneous force and cumulative work for the second model resemble those obtained experimentally on the unfolding of ARN molecules. Hence, the proposed systems are not just useful to illustrate the performance and conceptual significance of the fluctuation theorems, but also they could be studied as simplified models for biophysical systems.
\begin{description}
\item[Keywords]
Stochastic thermodynamics, Free energy difference, Nonequilibrium thermodynamics,\\
Fluctuation theorems, Molecular dynamics, Crooks Theorem.
\end{description}
\end{abstract}

\maketitle

\section{\label{sec:intro}Introduction}

Stochastic thermodynamics is a framework that enables the study of mesoscopic systems in nonequilibrium protocols where the interaction with the environment produces considerable fluctuations in the state of the system~\cite{Sekimoto}, for example, a particle undergoing Brownian motion. This approach has been successful in describing the behavior of biological systems at molecular level~\cite{bacterial_replisomes, CRISPR_Dynamics} and producing powerful theoretical tools as well as experimental procedures to measure relevant quantities such as free energy differences between molecule configurations~\cite{Sivak_protein_copy-number, Sivak_F1-ATPase_optimal_control}. 
Some of the most useful results within this theoretical framework are fluctuation theorems~\cite{Seifert_Principles,DetailedFT_Evans_Searles,Jarzynski1, Jarzynski2,Crooks,seifert_IFT}, that allow researchers to estimate free energy differences by gathering the work performed on many realizations of a nonequilibrium process. 

Fluctuation theorems like the Jarzynski equality~\cite{Jarzynski1, Jarzynski2}, the Crooks relation~\cite{Crooks} or the Hummer-Szabo relation~\cite{Hummer_Szabo} have become usual methods in biophysics to extract free energy differences by single-molecule pulling protocols or folding-unfolding protocols, both in experiments and simulations~\cite{bustamante_jarzynski, Bustamante_Crooks,severino_single-molecule_pulling_experiments} and in molecular motors~\cite{kinesin_crooks_FT,Fluctuation_Theorem_F1-ATPase}. Additionally, tools to implement folding-unfolding protocols and their analysis through the Crooks relation are included nowadays in high-performance molecular dynamics software for biophysics such as GROMACS~\cite{gromacs} and CHARMM~\cite{charmm}. 

Most folding-unfolding protocols at molecular level can be seen as the elongation and compression of masses joined by springs representing bonds that eventually can change, break or re-establish. 
Indeed, most molecular dynamics software packages model biomolecules as masses joined by springs~\cite{gromacs,charmm}. Moreover, the stochastic evolution of those models has analytical solution, so they can be implemented as reference systems for running benchmarking simulations. Thus, performing folding-unfolding protocols on simplified models of masses and springs will be useful to learn how fluctuation theorems, like the Crooks relation, perform and which ingredients must be taken into account to obtain reliable results. 

In this work we introduce two toy models of coupled harmonic oscillators under Brownian motion, and we perform simulations of fast-switching forward and reversed protocols on both systems, resembling the folding-unfolding protocols employed by Collin et al.~\cite{Bustamante_Crooks}. We use the Crooks fluctuation theorem to estimate the free energy difference between the initial and final states of the protocol, comparing those values with the theoretical free energy differences. The models are easy to solve and implement, and constitute excellent playgrounds to understand how fluctuation theorems perform to measure free energy differences.

\section{Theoretical Framework}

\subsection{Langevin Equation}

Let us consider a harmonic oscillator under the influence of a heat bath at temperature $T$. The basic formulation of such a system is the Langevin equation considering a mass $m$ under the effect of a random force, a viscous force, and a spring of constant $k$,
\begin{equation}\label{eq:langevin}
    m\ddot{x} = - \frac{\partial U(x,\lambda)}{\partial x} -m\gamma \dot{x}+bm\xi(t)\, ,
\end{equation}
where $x$ is the position of the particle, $U = \frac{k}{2}(x-x_c(t))^2$ is the harmonic oscillator potential that depends on the external parameter $x_c(t)$, $\gamma$ is the friction rate, and $\xi$ is a random noise, with $\langle \xi \rangle = 0$ and $\langle \xi(t)\xi(t')\rangle = \delta(t-t')$. Such equation describes the behavior of a particle undergoing Brownian motion with an additional external potential.  In this framework, the first law of thermodynamics is expressed as~\cite{Sekimoto}
\begin{align}\label{eq:first_law}
  dW &= dQ + dU\, ,
\end{align}
with
\begin{align}
    dQ &\equiv -\left(-m\gamma \dot{x} + bm\xi(t) \right)dx\, ,\\
    dW &\equiv \frac{\partial U}{\partial x_c}dx_c\, , \label{eq:work_definition}
\end{align}
where $dQ$ is the energy the system interchanges with the heat bath, and $dW$, the variation in potential energy due to the changes in the external parameter $x_c(t)$\footnote{Hereby, these definitions are expressed by means of Stratonovich calculus}.

By taking averages and solving the resultant ordinary differential equation \eqref{eq:langevin}, we can see that the oscillation is overdamped for $\gamma > 2\sqrt{\frac{k}{m}}$. In the overdamped limit, the position changes very slow. Therefore, we can approximate $\ddot{x} \sim 0$, and Eq. \eqref{eq:langevin} becomes
\begin{equation}
    \dot{x} = -\frac{k}{\gamma}(x-x_c)+\frac{b}{\gamma}\xi\, .
\end{equation}\\
This is the Fokker-Planck equation of an Ornstein-Uhlenbeck process, which can be solved by using the Kramers-Moyal expansion~\cite{Kramers, Moyal}. The result is that the position $x$ shows a normal distribution  $\rho(x)$ with mean $\mu_x = x_0\exp(-kt/m\gamma)$ and standard deviation $\sigma_x = \sqrt{\frac{k_\text{B}T}{k}(1-e^{-2kt/\gamma})}$~\cite{Ornstein_Uhlenbeck}.

\subsection{Free energy of coupled harmonic oscillators}

We also consider the exact free energy difference when coupling and uncoupling two harmonic oscillators. Let a mass $m$ coupled  with a spring of constant $k$ of zero natural length oscillate around an equilibrium position $x_c$. The partition function of such harmonic oscillator is
\begin{equation}
    Z(T) = \frac{1}{\beta \hbar \omega}\, ,
\end{equation}\\
where $\omega^2 = k/m$. Thus, its free energy is 
\begin{equation}
    F(T) = -k_\text{B}\ln Z(T) = k_\text{B}T \ln (\beta \hbar \omega).
\end{equation}

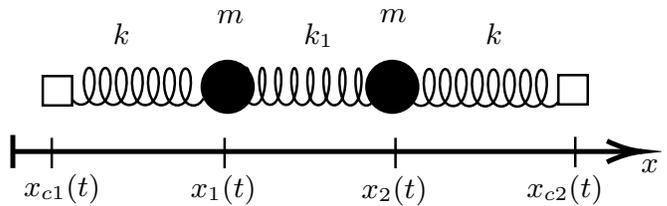
\begin{figure}[t]
    \centering
    \tikzset{every picture/.style={line width=0.75pt}} %set default line width to 0.75pt        

\begin{tikzpicture}[x=0.75pt,y=0.75pt,yscale=-1,xscale=1]

%Shape: Ellipse [id:dp6990609908464351] 
\draw  [fill={rgb, 255:red, 0; green, 0; blue, 0 }  ,fill opacity=1 ] (80.36,36.33) .. controls (80.36,30.54) and (85.05,25.85) .. (90.84,25.85) .. controls (96.62,25.85) and (101.31,30.54) .. (101.31,36.33) .. controls (101.31,42.11) and (96.62,46.8) .. (90.84,46.8) .. controls (85.05,46.8) and (80.36,42.11) .. (80.36,36.33) -- cycle ;
%Shape: Ellipse [id:dp7411006138210119] 
\draw  [fill={rgb, 255:red, 0; green, 0; blue, 0 }  ,fill opacity=1 ] (147.03,36.33) .. controls (147.03,30.54) and (151.72,25.85) .. (157.5,25.85) .. controls (163.29,25.85) and (167.98,30.54) .. (167.98,36.33) .. controls (167.98,42.11) and (163.29,46.8) .. (157.5,46.8) .. controls (151.72,46.8) and (147.03,42.11) .. (147.03,36.33) -- cycle ;
%Shape: Spring [id:dp5875863957396945] 
\draw  [color={rgb, 255:red, 0; green, 0; blue, 0 }  ,draw opacity=1 ] (81.01,41.04) .. controls (80.12,42.73) and (78.79,43.9) .. (76.95,43.88) .. controls (70.39,43.83) and (70.5,28.92) .. (73.81,28.94) .. controls (77.12,28.97) and (77.01,43.88) .. (70.45,43.83) .. controls (63.89,43.78) and (64,28.87) .. (67.31,28.89) .. controls (70.62,28.92) and (70.51,43.83) .. (63.95,43.78) .. controls (57.39,43.73) and (57.5,28.82) .. (60.81,28.84) .. controls (64.12,28.87) and (64.01,43.78) .. (57.45,43.74) .. controls (50.89,43.69) and (51,28.77) .. (54.31,28.79) .. controls (57.62,28.82) and (57.51,43.74) .. (50.95,43.69) .. controls (44.39,43.64) and (44.5,28.72) .. (47.81,28.75) .. controls (51.12,28.77) and (51.01,43.69) .. (44.45,43.64) .. controls (37.89,43.59) and (38,28.67) .. (41.31,28.7) .. controls (44.62,28.72) and (44.51,43.64) .. (37.95,43.59) .. controls (31.39,43.54) and (31.5,28.62) .. (34.81,28.65) .. controls (38.12,28.67) and (38.01,43.59) .. (31.45,43.54) .. controls (29.37,43.52) and (27.96,42.02) .. (27.11,39.97) ;
%Shape: Spring [id:dp5350730466093898] 
\draw  [color={rgb, 255:red, 0; green, 0; blue, 0 }  ,draw opacity=1 ] (150.32,40.94) .. controls (149.59,42.69) and (148.5,43.91) .. (147,43.9) .. controls (141.75,43.86) and (141.86,28.94) .. (143.86,28.96) .. controls (145.86,28.97) and (145.75,43.89) .. (140.5,43.85) .. controls (135.25,43.81) and (135.36,28.89) .. (137.36,28.91) .. controls (139.36,28.92) and (139.25,43.84) .. (134,43.8) .. controls (128.75,43.76) and (128.86,28.84) .. (130.86,28.86) .. controls (132.86,28.87) and (132.75,43.79) .. (127.5,43.75) .. controls (122.25,43.71) and (122.36,28.79) .. (124.36,28.81) .. controls (126.36,28.82) and (126.25,43.74) .. (121,43.7) .. controls (115.75,43.66) and (115.86,28.75) .. (117.86,28.76) .. controls (119.86,28.78) and (119.75,43.69) .. (114.5,43.65) .. controls (109.25,43.61) and (109.36,28.7) .. (111.36,28.71) .. controls (113.36,28.73) and (113.25,43.64) .. (108,43.6) .. controls (102.75,43.56) and (102.86,28.65) .. (104.86,28.66) .. controls (106.86,28.68) and (106.75,43.59) .. (101.5,43.56) .. controls (96.25,43.52) and (96.36,28.6) .. (98.36,28.61) .. controls (100.35,28.63) and (100.25,43.38) .. (95.08,43.51) ;
%Shape: Spring [id:dp4243056781350938] 
\draw  [color={rgb, 255:red, 0; green, 0; blue, 0 }  ,draw opacity=1 ] (224.59,42.52) .. controls (223.73,43.68) and (222.58,44.43) .. (221.11,44.41) .. controls (214.55,44.37) and (214.66,29.45) .. (217.97,29.47) .. controls (221.28,29.5) and (221.17,44.41) .. (214.61,44.37) .. controls (208.05,44.32) and (208.16,29.4) .. (211.47,29.42) .. controls (214.78,29.45) and (214.67,44.37) .. (208.11,44.32) .. controls (201.55,44.27) and (201.66,29.35) .. (204.97,29.38) .. controls (208.28,29.4) and (208.17,44.32) .. (201.61,44.27) .. controls (195.05,44.22) and (195.16,29.3) .. (198.47,29.33) .. controls (201.78,29.35) and (201.67,44.27) .. (195.11,44.22) .. controls (188.55,44.17) and (188.66,29.25) .. (191.97,29.28) .. controls (195.28,29.3) and (195.17,44.22) .. (188.61,44.17) .. controls (182.05,44.12) and (182.16,29.21) .. (185.47,29.23) .. controls (188.78,29.25) and (188.67,44.17) .. (182.11,44.12) .. controls (175.55,44.07) and (175.66,29.16) .. (178.97,29.18) .. controls (182.28,29.21) and (182.17,44.12) .. (175.61,44.07) .. controls (169.05,44.02) and (169.16,29.11) .. (172.47,29.13) .. controls (175.78,29.16) and (175.67,44.07) .. (169.11,44.02) .. controls (167.61,44.01) and (166.45,43.22) .. (165.61,42) ;
%Shape: Rectangle [id:dp38594101157402294] 
\draw  [fill={rgb, 255:red, 255; green, 255; blue, 255 }  ,fill opacity=1 ] (224.76,31.59) -- (236.64,31.59) -- (236.64,43.47) -- (224.76,43.47) -- cycle ;
%Straight Lines [id:da645903448690395] 
\draw [line width=1.5]    (3.84,62.56) -- (255.64,62.56) ;
\draw [shift={(258.64,62.56)}, rotate = 180] [color={rgb, 255:red, 0; green, 0; blue, 0 }  ][line width=1.5]    (14.21,-4.28) .. controls (9.04,-1.82) and (4.3,-0.39) .. (0,0) .. controls (4.3,0.39) and (9.04,1.82) .. (14.21,4.28)   ;
\draw [shift={(3.84,62.56)}, rotate = 180] [color={rgb, 255:red, 0; green, 0; blue, 0 }  ][line width=1.5]    (0,6.71) -- (0,-6.71)   ;
%Straight Lines [id:da6825757677065212] 
\draw    (89.6,57) -- (89.6,69.36) ;
%Straight Lines [id:da40708105351934254] 
\draw    (158.8,57) -- (158.8,69.36) ;
%Straight Lines [id:da4051156545407677] 
\draw    (19.6,57) -- (19.6,69.36) ;
%Straight Lines [id:da0876656285805899] 
\draw    (231.6,56.6) -- (231.6,68.96) ;
%Shape: Rectangle [id:dp10439292203804662] 
\draw  [fill={rgb, 255:red, 255; green, 255; blue, 255 }  ,fill opacity=1 ] (15.96,31.99) -- (27.84,31.99) -- (27.84,43.87) -- (15.96,43.87) -- cycle ;

% Text Node
\draw (84.95,3.7) node [anchor=north west][inner sep=0.75pt]    {$m$};
% Text Node
\draw (150.95,3.3) node [anchor=north west][inner sep=0.75pt]    {$m$};
% Text Node
\draw (256.8,64.2) node [anchor=north west][inner sep=0.75pt]    {$x$};
% Text Node
\draw (74.4,70.4) node [anchor=north west][inner sep=0.75pt]    {$x_{1}( t)$};
% Text Node
\draw (143.6,71) node [anchor=north west][inner sep=0.75pt]    {$x_{2}( t)$};
% Text Node
\draw (6.8,70) node [anchor=north west][inner sep=0.75pt]    {$x_{c1}( t)$};
% Text Node
\draw (210.4,70.4) node [anchor=north west][inner sep=0.75pt]    {$x_{c2}( t)$};
% Text Node
\draw (42.95,9) node [anchor=north west][inner sep=0.75pt]    {$k$};
% Text Node
\draw (193.75,9) node [anchor=north west][inner sep=0.75pt]    {$k$};
% Text Node
\draw (120.15,9) node [anchor=north west][inner sep=0.75pt]    {$k_{1}$};

\end{tikzpicture}
    \caption{Two harmonic oscillators of mass $m$, spring constant $k$, positions (black circles) $x_1(t)$ and $x_2(t)$, and coupled by a third spring with constant $k_1$. The natural positions of each individual spring $x_{c1}(t)$ and $x_{c2}(t)$ are represented with white squares.}
    \label{fig:Coupled_springs}
\end{figure}

Now, let us consider two of such harmonic oscillators of mass $m$ and spring constant $k$, either uncoupled or coupled by a third spring of constant $k_1$ and natural length $2d_1$ (Fig. \ref{fig:Coupled_springs}).
On the one hand, two uncoupled harmonic oscillators are just two non-interacting systems, and its total free energy $F_u$ is just twice the free energy of a single one, $F_u=2F(T)$. On the other hand, when the two masses are coupled by the third spring $k_1$ we can diagonalize the system and consider each normal mode as a non-interacting harmonic oscillator. For two oscillators with frequency $\omega$, the normal modes are
\begin{equation}
    \omega_1^2 = \frac{k}{m}\, ,\quad \omega_2^2 = \frac{k + 2k_1}{m}\, .
\end{equation}
Thus, the free energy $F_c$ is
\begin{equation}
    F_c = k_\text{B}T \ln (\beta^2\hbar^2\omega_1\omega_2)\, ,
\end{equation}
Nevertheless, this expression assumes that the coupled system is in an equilibrium configuration where all springs are elongated their natural lengths and the elastic potential energy is $U_0=0$, but this is not always the case. Let us consider that, when uncoupled, the two masses oscillate around equilibrium positions $x_{c1}=-d_0$ and $x_{c2}=d_0$, with $d_0<d_1$. When coupled, the spring $k_1$ is compressed, and the equilibrium positions for the two masses are $x_{eq2}=(kd_0+2k_1d_1)/(k+2k_1)$ and $x_{eq1}=-x_{eq2}$, respectively. The potential energy of such configuration is
\begin{equation}\label{eq:potential_energy}
    U_0=\frac{2}{9} (k+2k_1) (d_1-d_0)^2 \, ,
\end{equation}
and the free energy for the coupled system becomes
\begin{equation}
    F_c = k_\text{B}T \ln (\beta^2\hbar^2\omega_1\omega_2)+U_0\, .
\end{equation}
Thus, the free energy change when uncoupling two oscillators is
\begin{equation}\label{eq:F_ens}
    \Delta F_\text{ens} = F_u - F_c = k_\text{B}T\ln \left(\frac{\omega^2}{\omega_1\omega_2}\right)-U_0\, .
\end{equation}

\subsection{Crooks Fluctuation Theorem}

Fluctuation theorems have shown researchers how to gather information of equilibrium states from measurements taken through nonequilibrium processes. Particularly, the Crooks fluctuation theorem (CFT), proposed by Gavin Crooks~\cite{Crooks} in 1999, relates the probability that a system produces a certain amount of work $\Delta W$ when undergoing a specific protocol at constant temperature $T$ with the probability that it produces a work $-\Delta W$ when going through the reverse protocol
\begin{equation}\label{eq:crooks}
    P_\text{for}(\Delta W) = \exp\left(\frac{\Delta W-\Delta F}{k_\text{B}T}\right) P_\text{rev}(-\Delta W)\, .
\end{equation}
By plotting $\log\left[P_\text{for}(\Delta W)/P_\text{rev}(-\Delta W)\right]$ vs. $\Delta W$, we estimate $\Delta F$ as the point where the curve crosses the horizontal axis (as shown in Figs. \ref{fig:Crooks_system1} and \ref{fig:Crooks_system2}).

\section{\label{sec:comp}Computational Methods}

In order to simulate a finite temperature on the system, we use the stochastic leapfrog thermostat algorithm for molecular dynamics~\cite{Gunsteren_Berendsen1}. This algorithm introduces a stochastic impulse to velocity $\Delta v$ in the classic leapfrog scheme to ensure that the temperature remains constant. A step in the algorithm for a single particle is as follows:
\begin{align}
    &v' = v_{t-\frac{\Delta t}{2}} + \frac{F}{m}\Delta t\, ,\\
    &\Delta v = -\alpha v' + \sqrt{\alpha(2-\alpha)(k_\text{B} T_{\text{ref}}/m)}\xi\, ,\label{eq:impulsive_step}\\
    &x_{t+\Delta t} = x_t + \left(v' + \frac{1}{2}\Delta v\right)\Delta t\, ,\\
    &v_{t+\frac{\Delta t}{2}}=v'+\Delta v\, ,
\end{align}\\
where $F$ is the deterministic force acting upon a particle of mass $m$, $T_\text{ref}$ is the constant temperature, and $\xi$ is a random variable with a normal distribution of mean $\langle \xi \rangle = 0$ and variance $\sigma_x^2=1$. 
The factor $\alpha$ is called the \textit{impulsive friction}, with $0\leq \alpha \leq 1$, and is related with $\gamma$ (Eq. \eqref{eq:langevin}) as
\begin{equation}
    \alpha = 1 - e^{-\gamma \Delta t}.
\end{equation}
In the absence of random and potential forces, the particle's velocity would be reduced at each time step by a factor $1-\alpha$.
The stationary distribution of the velocity in the absence of external forces is the Maxwell-Boltzmann distribution. It is important to mention that this algorithm does not conserve the momentum nor the energy of the system because of the random term in the velocity; however, this is the expected behavior from Eq. \eqref{eq:langevin}.

\section{\label{sec:models}The models}

The first system consists of two identical harmonic oscillators of mass $m$ and spring constant $k_0$ oscillating with positions $x_1$ and $x_2$ around centers $x_{c1}$ and $x_{c2}$, respectively.  The natural length of those springs is assumed equal to zero. When coupled, the two oscillators are joined by a third spring with the same constant $k_0$ and a natural length $2d_1$. Along with the spring forces, the two masses undergo a Brownian motion induced by the random and frictional forces from a surrounding medium at temperature $T$, as described in Eq. \eqref{eq:langevin}, but in the overdamped regime ($\gamma > 2\sqrt{3k_0/m}$).

\begin{figure}[b]
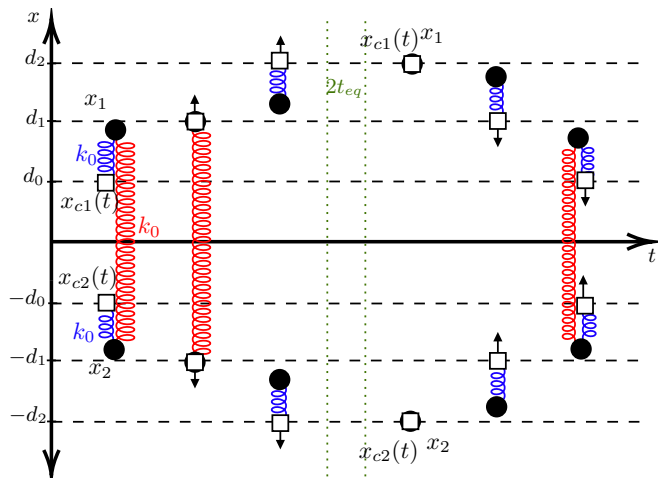

    \centering
    \includestandalone[width=0.49\textwidth]{Figures/diagram1}
    \caption{Forward (uncoupling) and time-reversed (coupling) protocols for the first  model of two coupled harmonic oscillators. The masses are represented by black circles, and the centers of each oscillator by white squares. The springs of each oscillator are shown in blue, and the coupling spring in red.}
    \label{fig:Diagram_system1}
\end{figure}

The system undergoes the following forward protocol: The two oscillators start at $x_{c2}=-x_{c1}=d_0$,  coupled by the third spring. Initially, we let the system to evolve for a time $t_\text{eq}=100/\gamma$ to reach equilibrium. Then, we start to pull away the centers of each oscillator at constant speed $v$ ($x_{c2}(t)=d_0+vt$, $x_{c1}(t)=-x_{c2}(t)$). When the distance between the masses reaches the natural length of the coupling spring (that is, when $x_2-x_1=2d_1$) it breaks, leaving the masses uncoupled. We keep pulling the center of each mass until they reach a final position $x_{c2}(t_f)=-x_{c1}(t_f)=d_2$. Then, we let the system relax a time $2t_\text{eq}$ before performing the reverse protocol, which consists in pushing the oscillators centers $x_{c1}$, $x_{c2}$ back together at the same speed we used to separate them. When the distance between the masses reaches the natural length of the coupling spring ($x_2-x_1=2d_1$) the spring reappears, coupling the masses again. A scheme of the forward and reverse protocols can be seen in Fig. \ref{fig:Diagram_system1}. 

Because all three springs have the same constant ($k=k_1=k_0$ in Eq. \eqref{eq:potential_energy}), the potential energy for the coupled system is $U_0=\frac{2}{3} k_0 (d_1-d_0)^2$, and the free energy difference between initial and final canonical ensembles in the forward protocol is 
\begin{equation}\label{eq:F_system1}
    \Delta F_{tot} = k_\text{B}T \ln \left(\frac{w^2}{w_1w_2}\right) -\frac{2}{3}k_0(d_1 - d_0)^2\, .
\end{equation}

\begin{figure}[hb!]
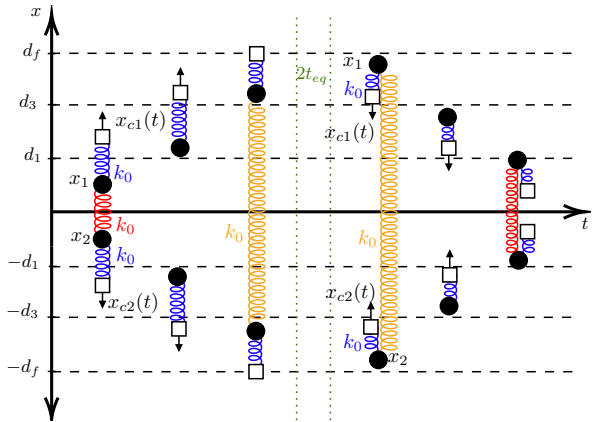

    \centering
    \includestandalone[width=0.44\textwidth]{Figures/diagram2}
    \caption{Diagram of the second system. Different steps in the forward and time-reversed protocols are shown. The second coupling spring is shown in orange.}
    \label{fig:Diagram_system2}
\end{figure}

Next, we considered a second model inspired by the force-displacement curves obtained by Collin et al.~\cite{Bustamante_Crooks}. Let us start from the first model and add a second coupling zone as follows: 
Once the two oscillators have uncoupled and the distance between the masses surpasses the value $2d_3$, we add a new coupling spring with natural length $2d_3$ and spring constant $k_0$ (Fig. \ref{fig:Diagram_system2}). Next, we keep pulling the oscillators' centers apart at a constant speed until reaching final positions $x_{c2}=d_f$, with $d_f=2d_1$ and wait a time $2t_\text{eq}$ for the system to relax. Then, we perform the reverse protocol by closing the centers at a constant speed until reaching their initial positions. When the distance between the masses equals $2d_3$, the new coupling spring disappears; when that distance equals $2d_1$, the first coupling spring reappears, as before. A scheme of this protocol can be seen in Fig \ref{fig:Diagram_system2}.

In this second model the system starts and ends in coupled states. Therefore, the free energy difference is just the difference between initial $U_{0i}=\frac{2k_0}{3}(d_1 - d_0)^2$ and final $U_{0f}=\frac{2k_0}{3}[(d_3-d_f)^2$ potential energies, 
\begin{equation}
    \Delta F_{tot} = \frac{2k_0}{3}[(d_3-d_f)^2-(d_1 - d_0)^2]\, .
\end{equation}

\section{Implementation and results}

Now, we want to verify the Crooks Fluctuation Theorem by simulating the two models with the protocols described in the previous section. For all cases we choose masses $m=1$, spring constants $k_0=4$, and a temperature $k_\text{B}T=4$. The friction rate must assure that the system is overdamped, but the relaxation time is preferably short; so we choose 
\begin{equation}
    \gamma = 6.92821 > 2\sqrt{3k_0/m}\, ,
\end{equation}
which exceeds the overdamped condition for the faster normal mode.

For the first model, the distances are chosen as $d_0=\pm 3$, $d_1=\pm 6$, and $d_2=\pm 9$ and, therefore, the initial equilibrium position for the masses is $x_\text{eq}=\pm 5$. When the oscillators centers are pulled apart or pushed together with speeds equal to or less than $v=\frac{10}{48}$, the time step is set at $\Delta t = 0.01$, and it is set to $\Delta t=0.001$ for higher speeds. The relaxation time is $t_\text{eq}=100$, which is one order of magnitude larger than $\tau=2m/\gamma$, the relaxation time constant for that system. By using Eq. \eqref{eq:F_system1}, the free energy difference for one particle is $\frac{\Delta F_\text{tot}}{2} = -13.0986$.

To verify that our simulation is well implemented, we estimated $\Delta F_\text{tot}$ by executing the cycle of one forward and one reverse protocol with a speed $v_\text{eq}=\frac{6}{14800}$, which is slow enough to consider a quasi-static process. The total time to perform the whole process is $t_\text{n}=30000$. 
According to Eq. \eqref{eq:work_definition}, the external work $W_{\rm ext}$ on one of the two masses along $n$ time steps is computed by integrating the force $F_0(t)$ exerted by the spring connected to the oscillator's center at $x_c$ (blue springs in Figs. \ref{fig:Diagram_system1} and \ref{fig:Diagram_system2}) times the small displacements of that center,
\begin{equation}
    W_{\rm ext} = \sum_{i=1}^{n} F_0(t_i)(x_c(t_{i})-x_c(t_{i-1}))\, ,
\end{equation}
where $t_n$ is the time the process ends.

\begin{figure}[h]
    \centering
    \includegraphics[width=\linewidth]{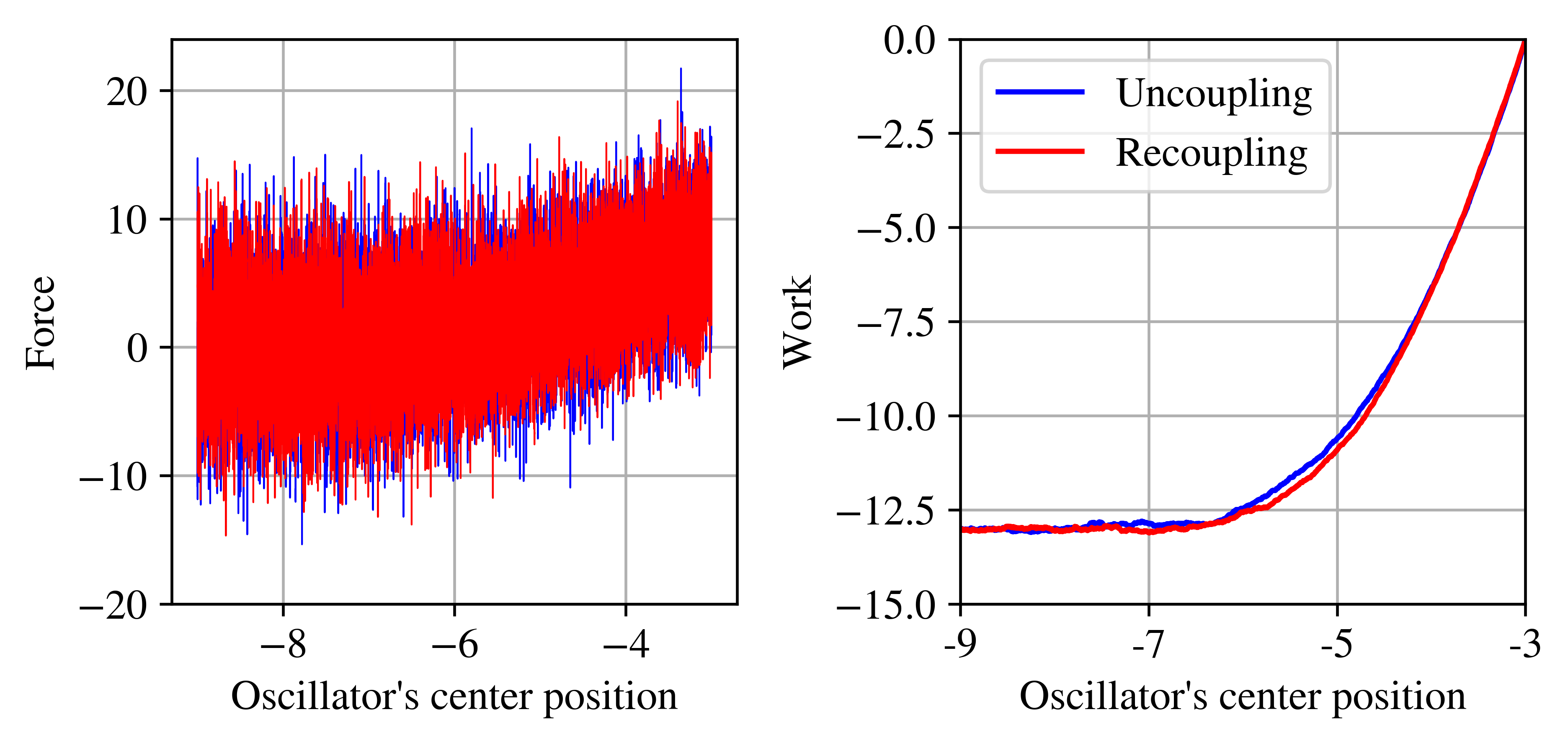}
    \caption{a. Instantaneous force and b. work exerted on one particle in a single trajectory at $v=\frac{6}{14800}$ for the first model. The uncoupling protocol is shown in blue and the coupling one, in red.}
    \label{fig:system1_work-force_single_trajectory}
\end{figure}

Figure \ref{fig:system1_work-force_single_trajectory} shows the instantaneous force exerted by a blue spring against $x_c$ for particle 2 on a single trajectory, together with the work performed on that mass, for both the forward (blue) and backward (red) protocols. 
It is noticeable that the work done in the forward protocol is nearly the same as the negative work done in the backward protocol, even for a single cycle. 

%AQUI VAMOS

\begin{figure}[h]
    \centering
    \begin{subfigure}{0.46\columnwidth}
        \centering
         \includegraphics[width=\linewidth]{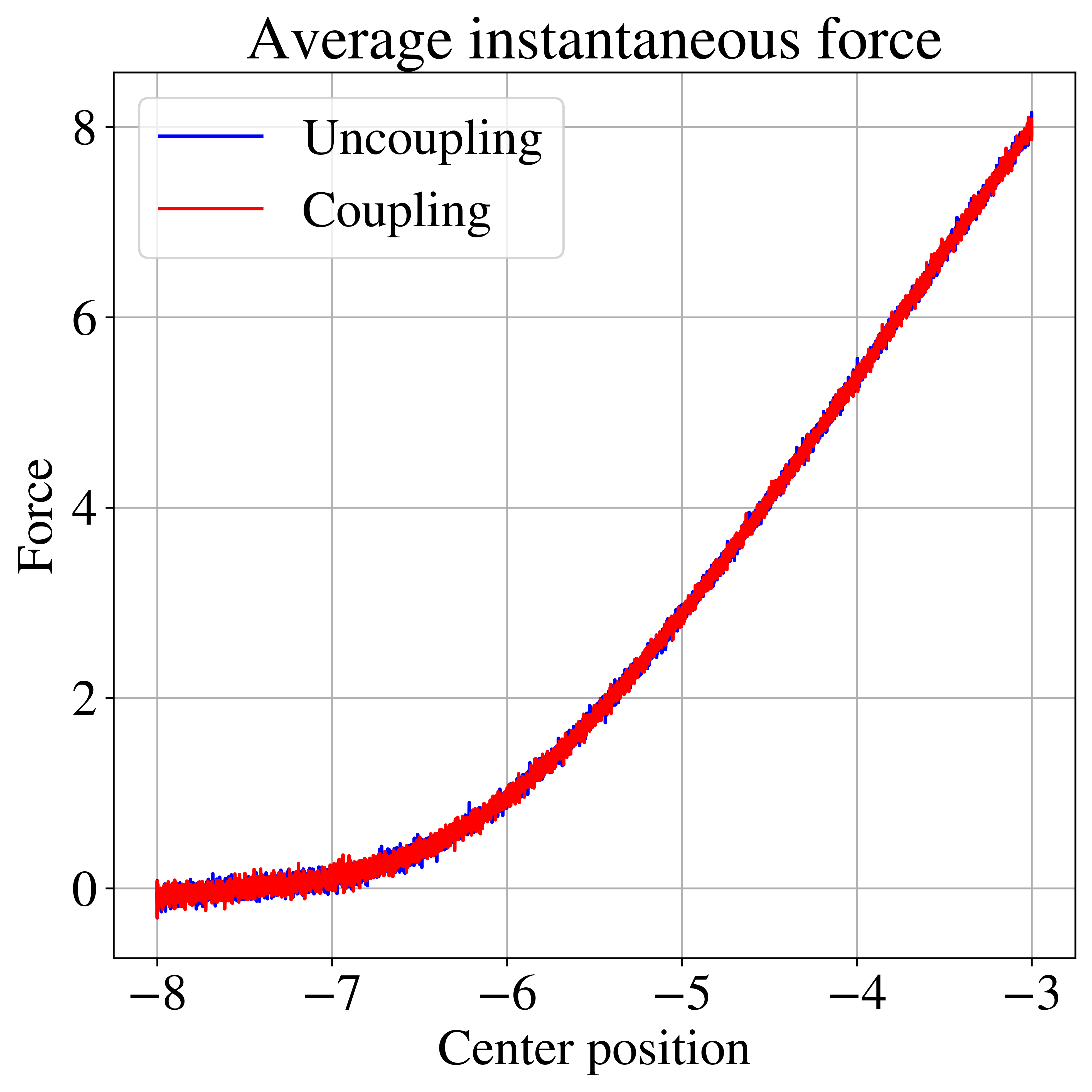}
    \end{subfigure}
    \hfill
    \begin{subfigure}{0.46\columnwidth}
        \centering
         \includegraphics[width=\linewidth]{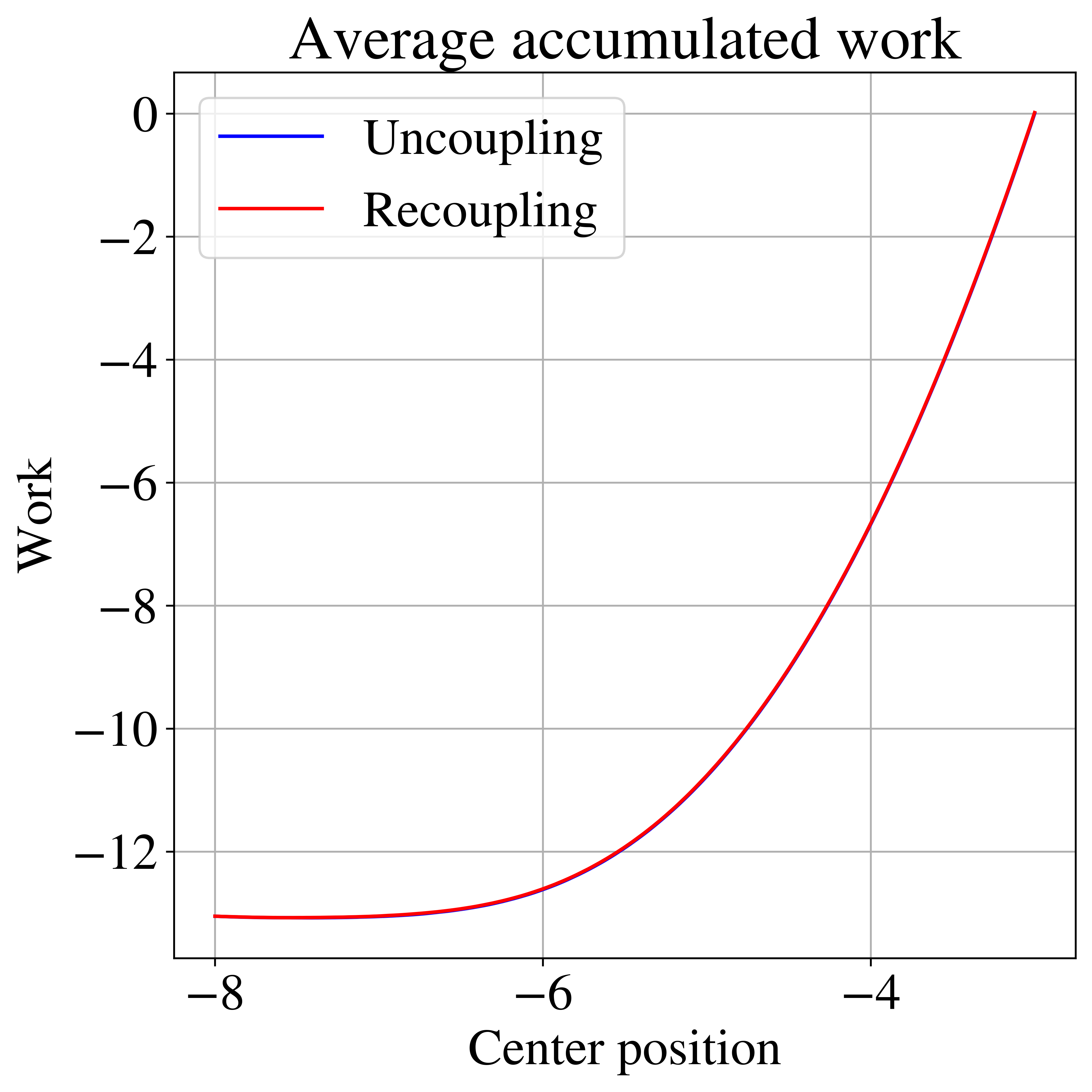}
    \end{subfigure}
    \caption{Averaged instantaneous force and work for one of the particles of the first model at $v=\frac{6}{14800}$.}
    \label{fig:system1_work-force}
\end{figure}

This observation can be reaffirmed by taking averages on  1000 trajectories (Fig. \ref{fig:system1_work-force}). For the averaged work, the maximal difference between the forward and reverse protocols is $0.008$, supporting our assumption of a quasi-static process.
The total free energy difference is the work performed on both masses,
\begin{equation}
    \frac{\Delta F_\text{tot}}{2} = -13.0926\, ,
\end{equation}
which differs from the theoretical value by $0.05\%$.

\begin{figure}[h]
    \centering
    \begin{subfigure}{0.46\columnwidth}
        \centering
         \includegraphics[width=\linewidth]{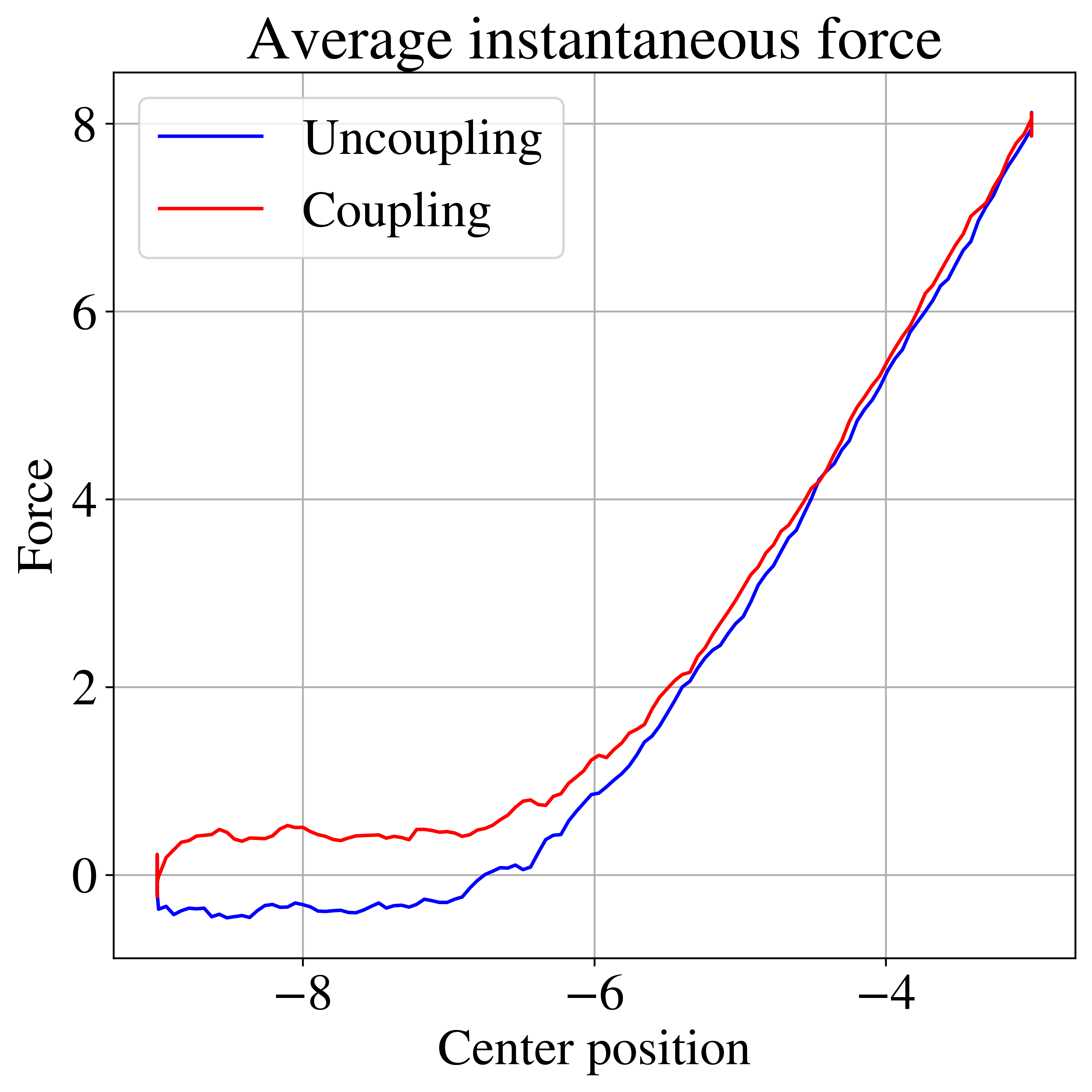}
    \end{subfigure}
    \hfill
    \begin{subfigure}{0.46\columnwidth}
        \centering
         \includegraphics[width=\linewidth]{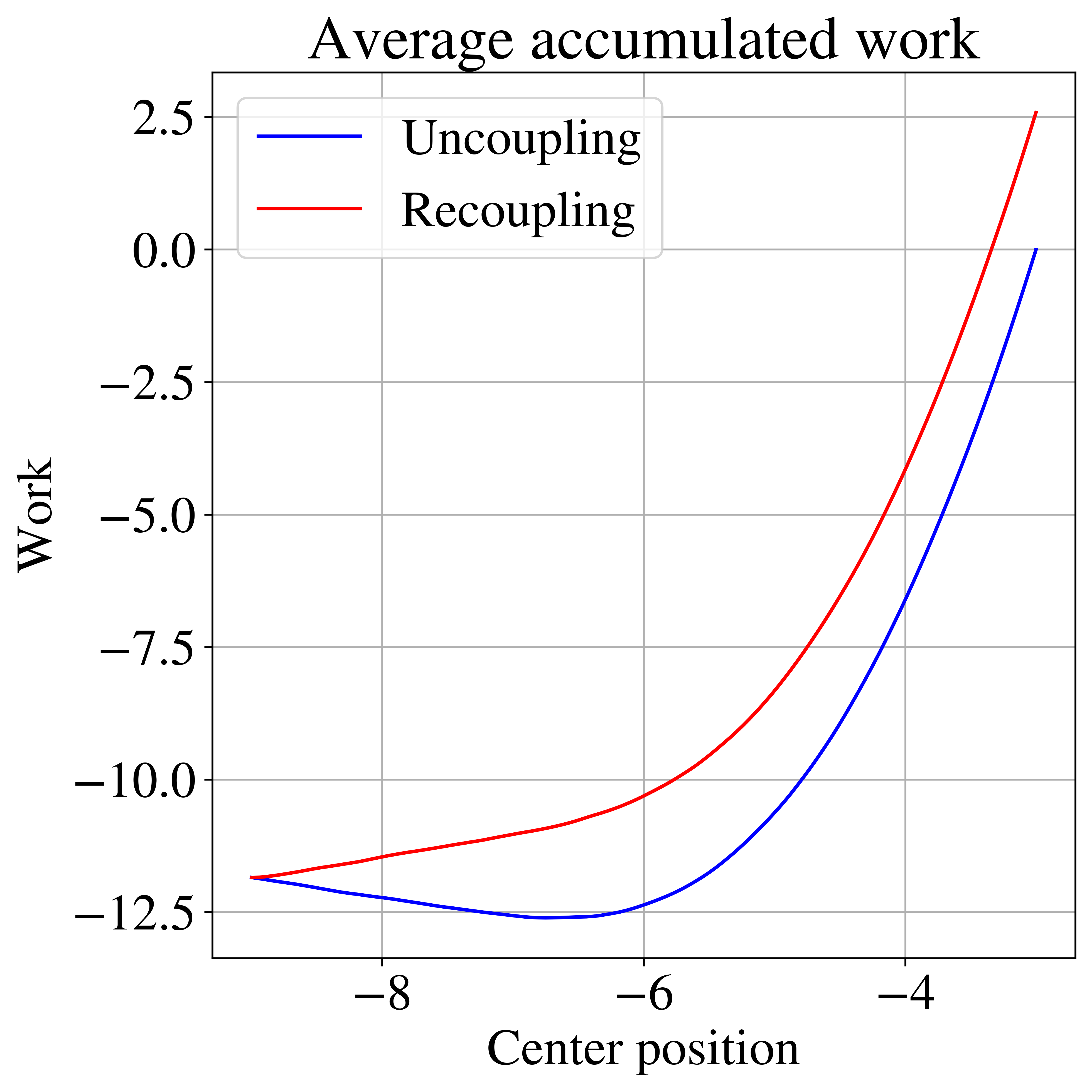}
    \end{subfigure}
    \caption{Averaged instantaneous force and work for one of the particles of the first model at $v=\frac{5}{48}$.}
    \label{fig:system1_work-force_v5_48}
\end{figure}

To test the Crooks fluctuation theorem we repeat the process with higher speeds. Figure \ref{fig:system1_work-force_v5_48} shows the average instantaneous force and average work for the process at velocity $v=\frac{5}{48}$. It can be seen that the work done in the coupling and uncoupling protocols are not the same, and the instantaneous force shows hysteresis. We repeat this treatment for seven speeds up to three orders of magnitude greater than the one used for the quasi-static process.

\begin{figure}[h!]
    \centering
    \includegraphics[width=\linewidth]{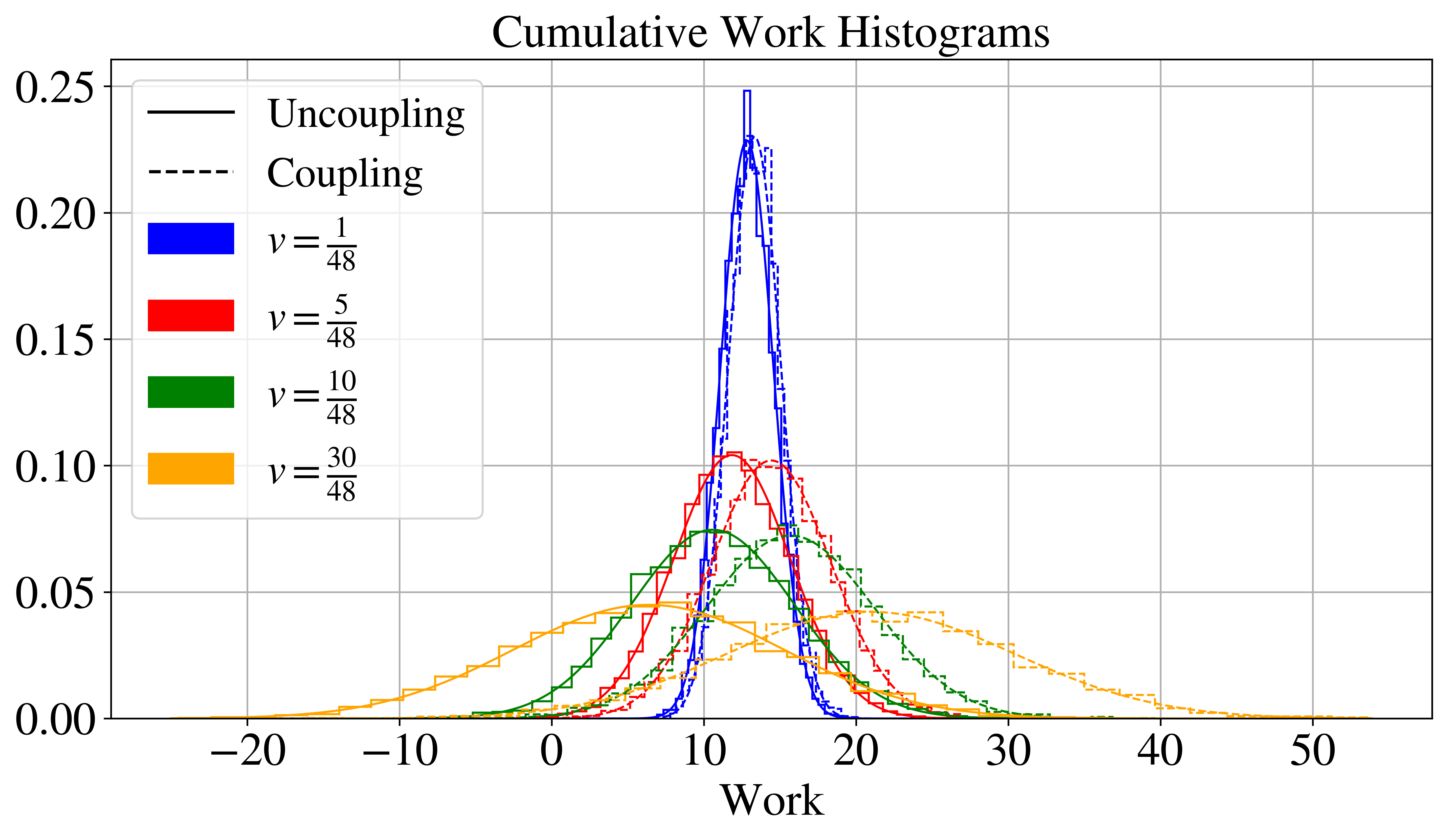}
    \caption{Histograms of work done for different velocities in the first model. Work done in the uncoupling and coupling protocols are shown in continuous and dashed lines respectively. Gaussian fits are superposed on each histogram.}
    \label{fig:work_distribution_system1}
\end{figure}

 Figure \ref{fig:work_distribution_system1} shows the histograms for the works done at each speed. It is noticeable that, when speed increases, the histograms for forward and reverse protocols distance from each other.

According to Eq. \eqref{eq:crooks}, $\Delta W = \Delta F$ at the point where $P_\text{for}(\Delta W)=P_\text{rev}(-\Delta W)$, that is when the histograms cross each other. From Fig. \ref{fig:work_distribution_system1}, we obtain $\Delta F_\text{tot} \simeq -13.37 \pm 0.24$, which differs from the theoretical value in just a 2.1\%.

\begin{figure}[h]
    \centering
    \includegraphics[width=\linewidth]{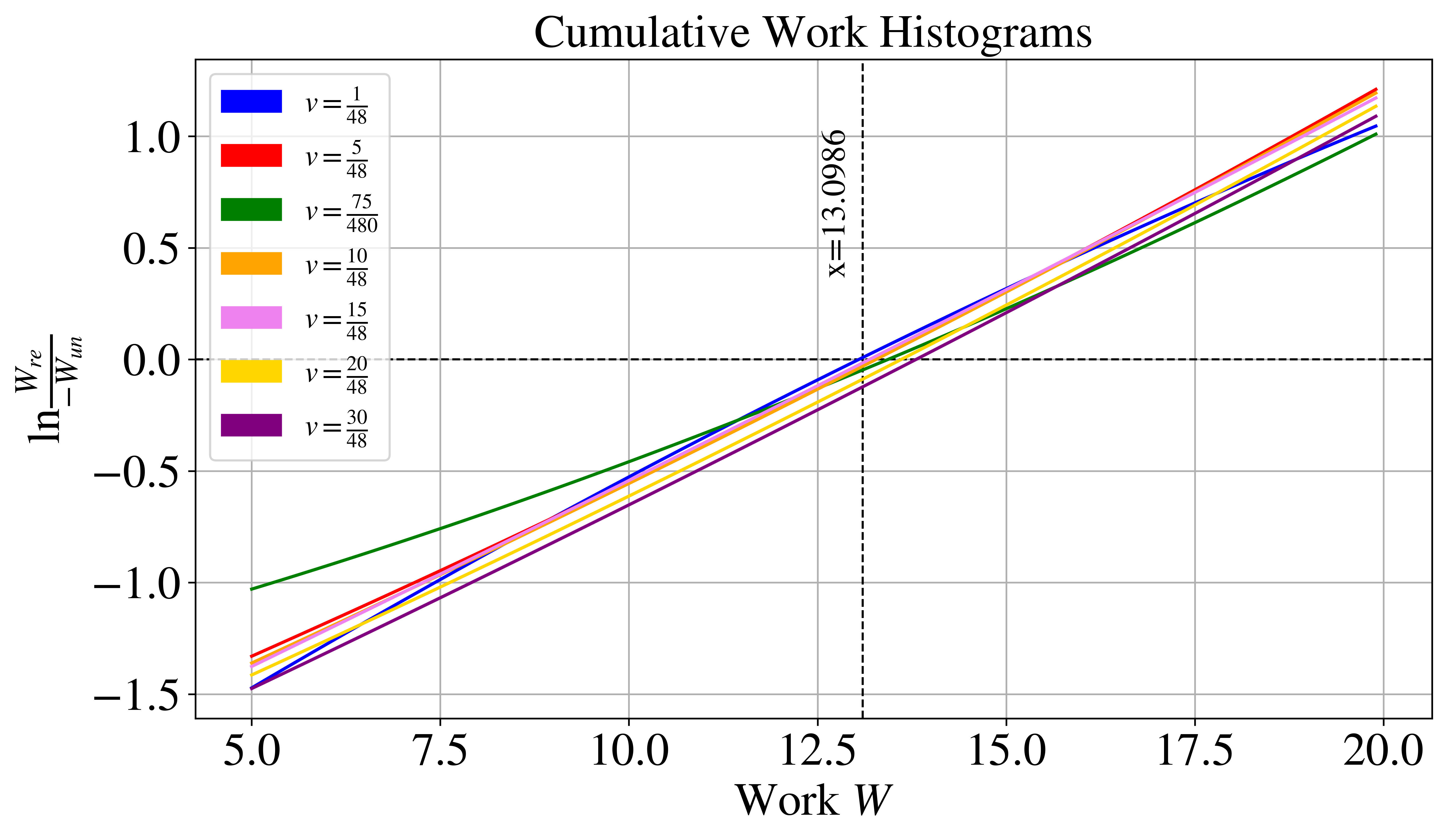}
    \caption{Crooks theorem for many velocities in the first model.}
    \label{fig:Crooks_system1}
\end{figure}

The free energy difference can also be computed by plotting $\ln \frac{P_\text{for}}{P_\text{rev}}$ against $\Delta W$, which is expected to be a straight line. The value $\Delta W$ where the line intersects the horizontal axis gives $\Delta F$. Fig. \ref{fig:Crooks_system1} verifies that linear relationship for each speed; moreover, all lines intersect the horizontal axis around the same value. By taking average over those values we obtain a free energy difference of $\Delta F_\text{tot}\simeq -13.26 \pm 0.32$ which differs in just $1.2\%$ from the theoretical prediction.

For the second model, we choose the same masses and spring constants $m=1$ $k=k_0=4$ and distances $d_0=3$, $d_1=6$, $d_3=9$, and $d_f=12$; thus, the exact free energy difference is null, $\Delta F_\text{tot}=0$. 

\begin{figure}[h]
    \centering
    \begin{subfigure}{0.46\columnwidth}
        \centering
         \includegraphics[width=\linewidth]{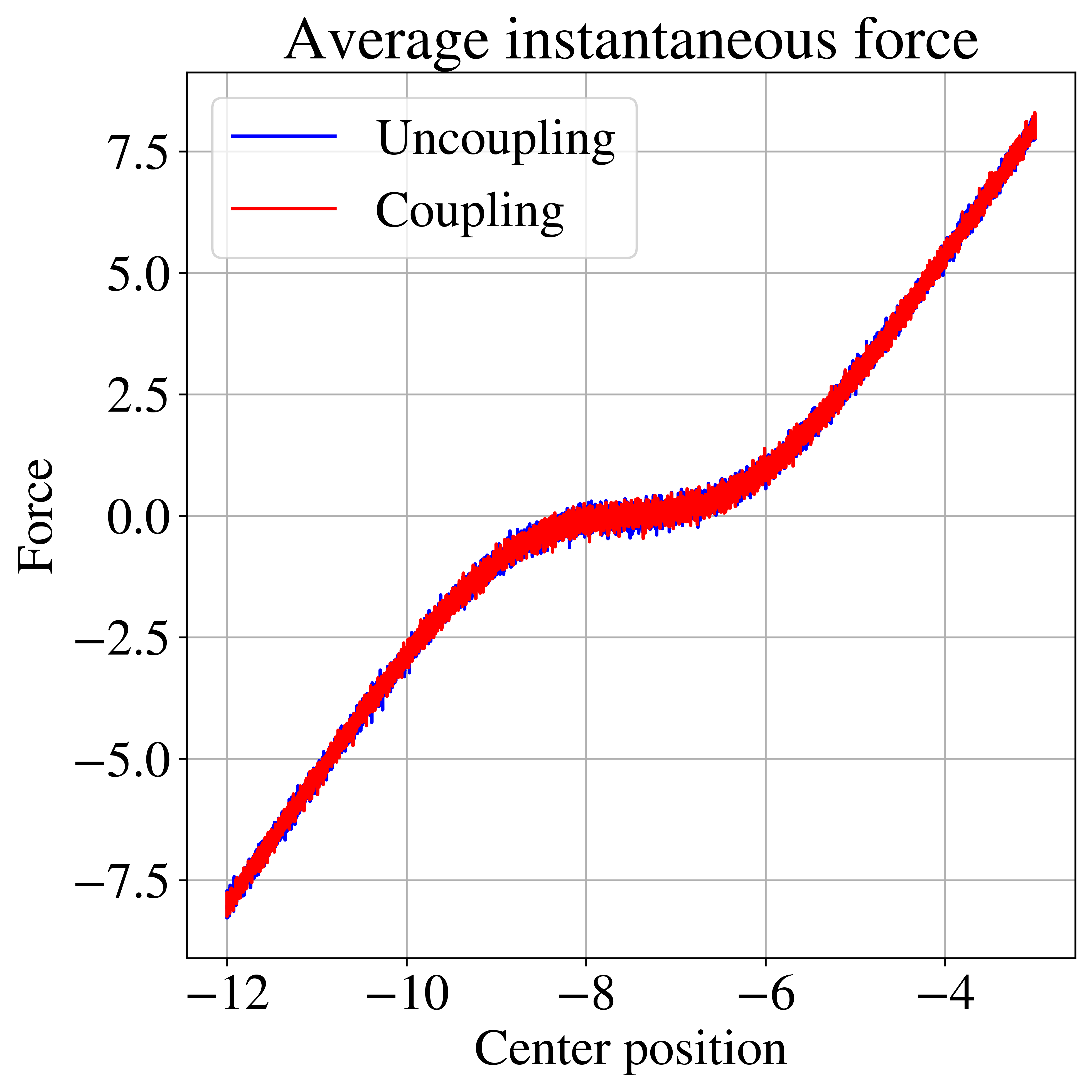}
    \end{subfigure}
    \hfill
    \begin{subfigure}{0.46\columnwidth}
        \centering
         \includegraphics[width=\linewidth]{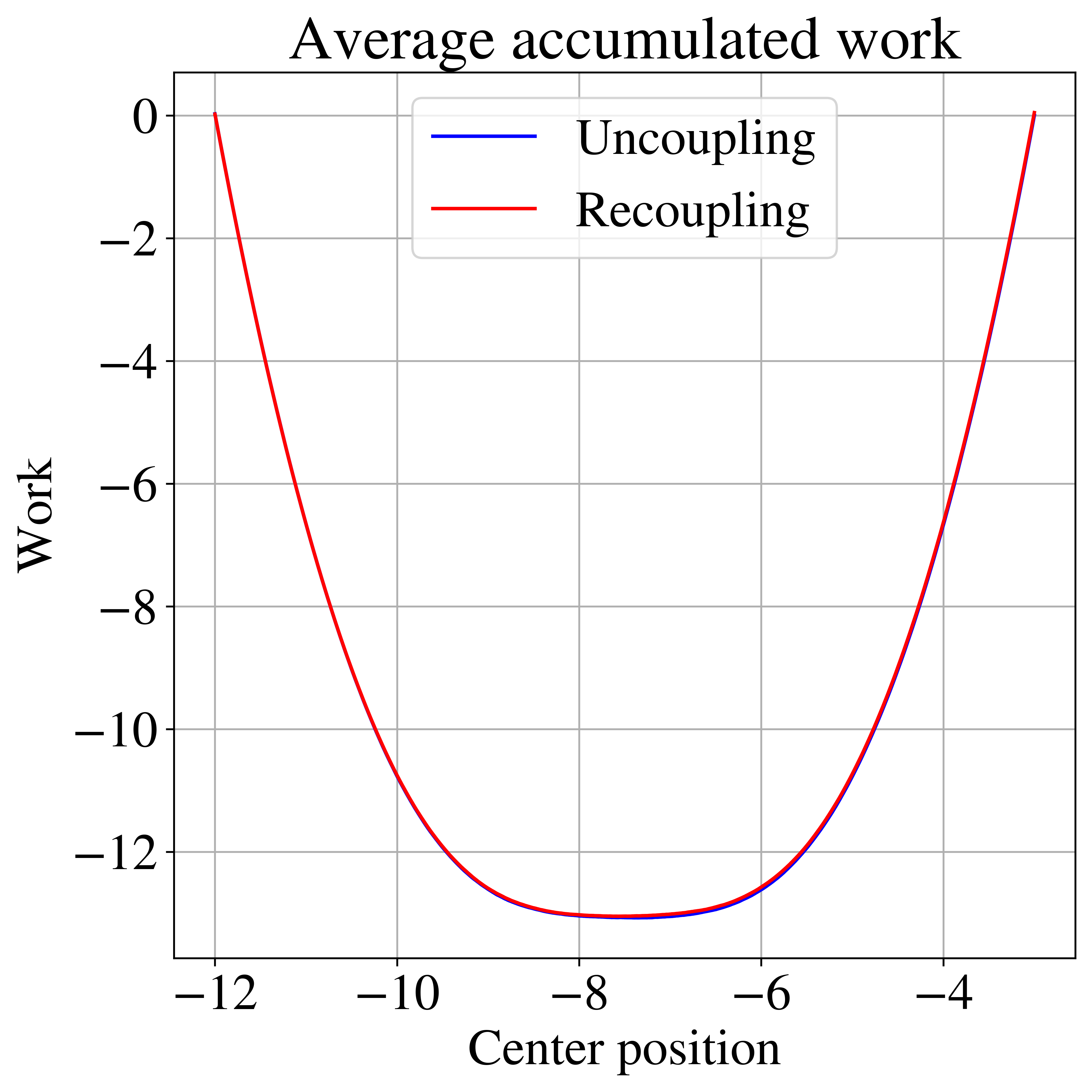}
    \end{subfigure}
    \caption{Averaged instantaneous force and work for one of the particles of the second model at $v=\frac{9}{14800}$.}
    \label{fig:system2_work-force}
\end{figure}

For this second model we chose a speed $v=\frac{9}{14800}$ to reproduce a quasi-static process. Figure \ref{fig:system2_work-force} shows the average instantaneous force and work for this second system. Once again, we observe that the coupling and uncoupling work are nearly the same, with a maximal difference between the forward and backward curves of $\Delta W = 0.02$. The free energy difference estimated from this quasi-static process is $\Delta F_\text{tot}=0.03\pm 0.04$, which is consistent with the exact result. 

\begin{figure}[t]
    \centering
    \begin{subfigure}{0.46\columnwidth}
        \centering
         \includegraphics[width=\linewidth]{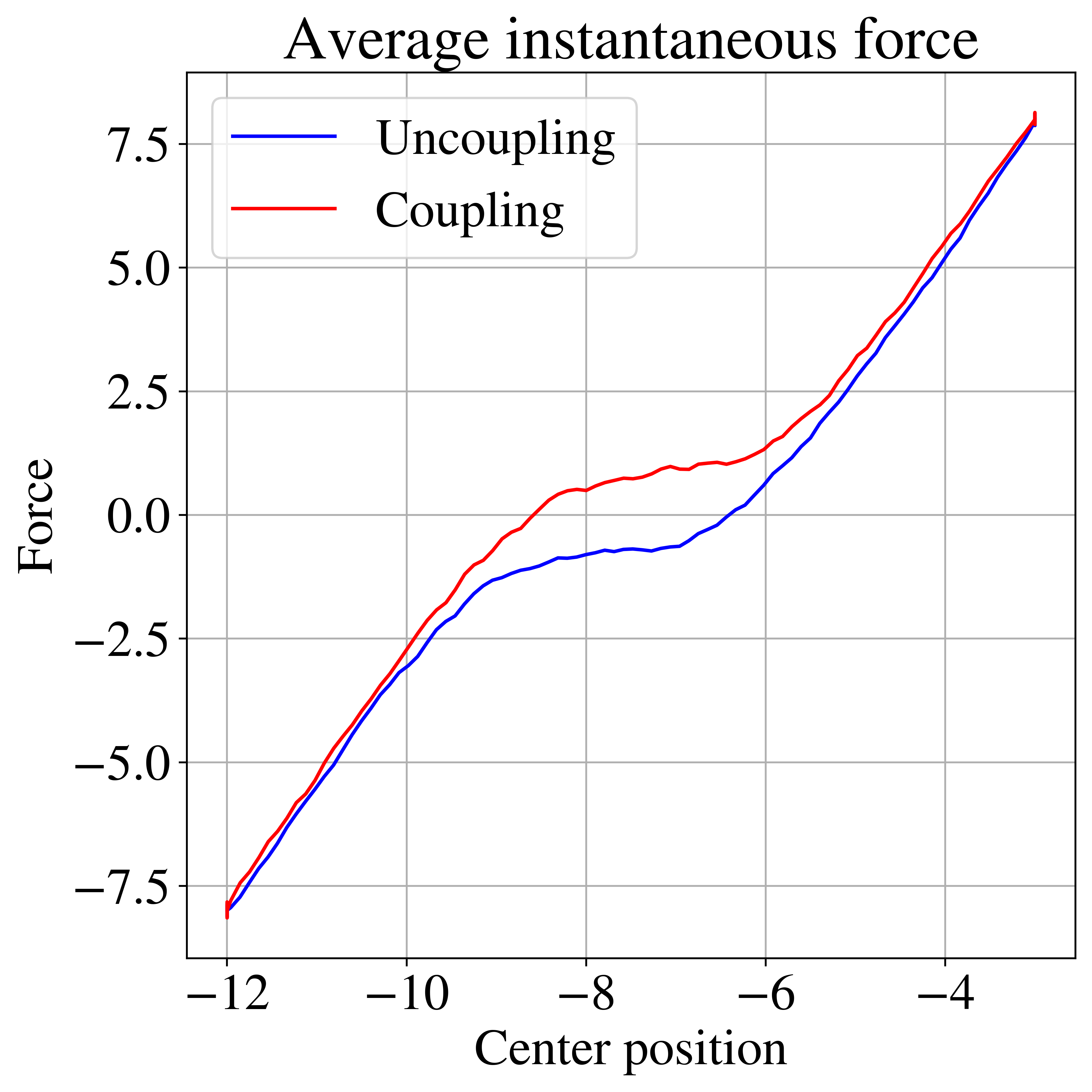}
    \end{subfigure}
    \hfill
    \begin{subfigure}{0.46\columnwidth}
        \centering
         \includegraphics[width=\linewidth]{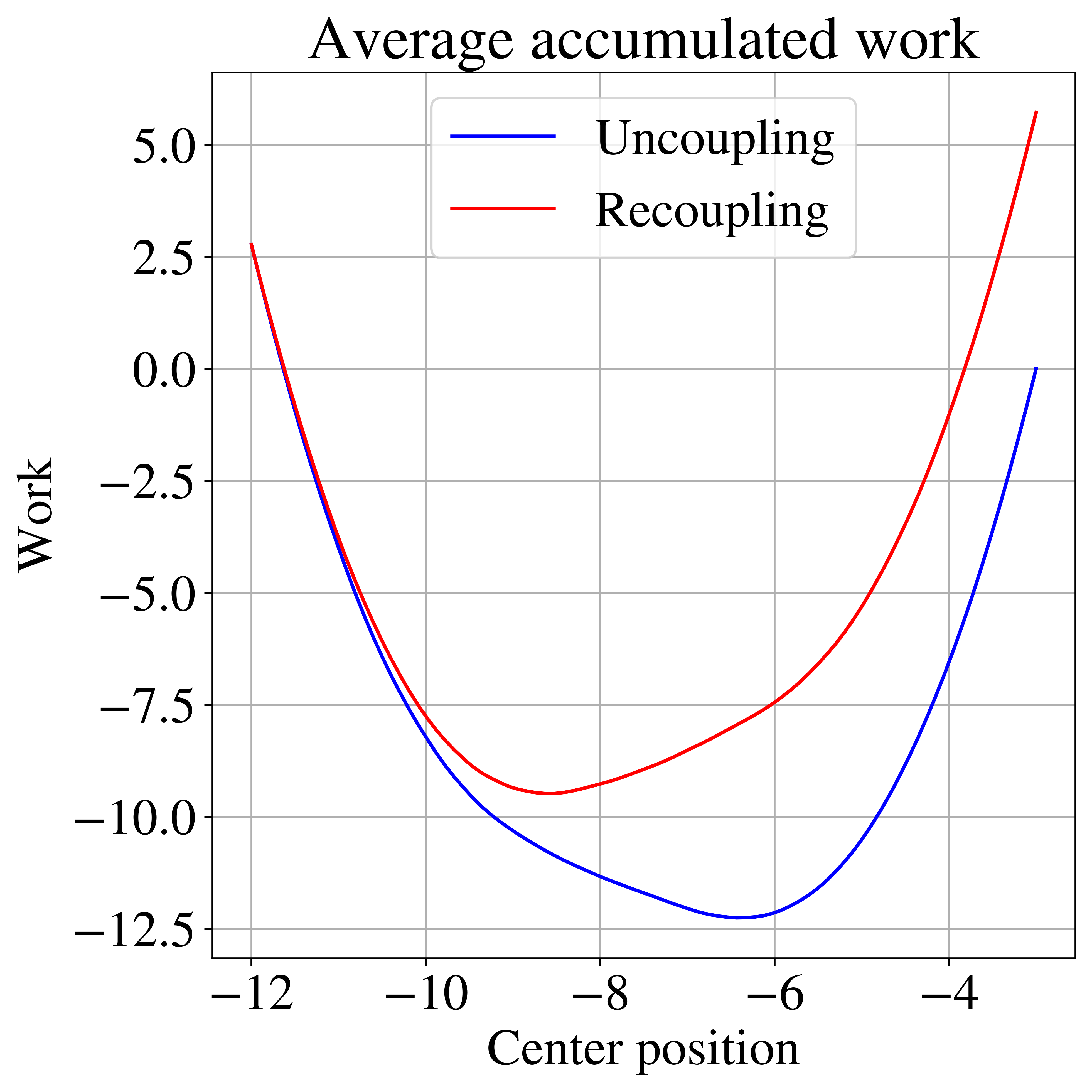}
    \end{subfigure}
    \caption{Averaged instantaneous force and work for one of the particles of the first model at $v=\frac{10}{48}$.}
    \label{fig:system2_work-force_v10_48}
\end{figure}

Figure \ref{fig:system2_work-force_v10_48}, shows the average force and work obtained when the protocol is performed at a speed of $v=\frac{10}{48}$. Again, we can see hysteresis. It is interesting to note that the graph for the force is similar to that obtained in an experimental setup by Collin et al.~\cite{Bustamante_Crooks}. Indeed, in that experiment the authors pull a molecule until it unfolds, which produces a momentary diminution in the force needed to pull it, before it tightens again. That mechanical behavior resembles the one of a spring that breaks. Thus, it is coherent that our simplified model displays the same behavior than the experiment.

\begin{figure}[h]
    \centering
    \includegraphics[width=0.42\textwidth]{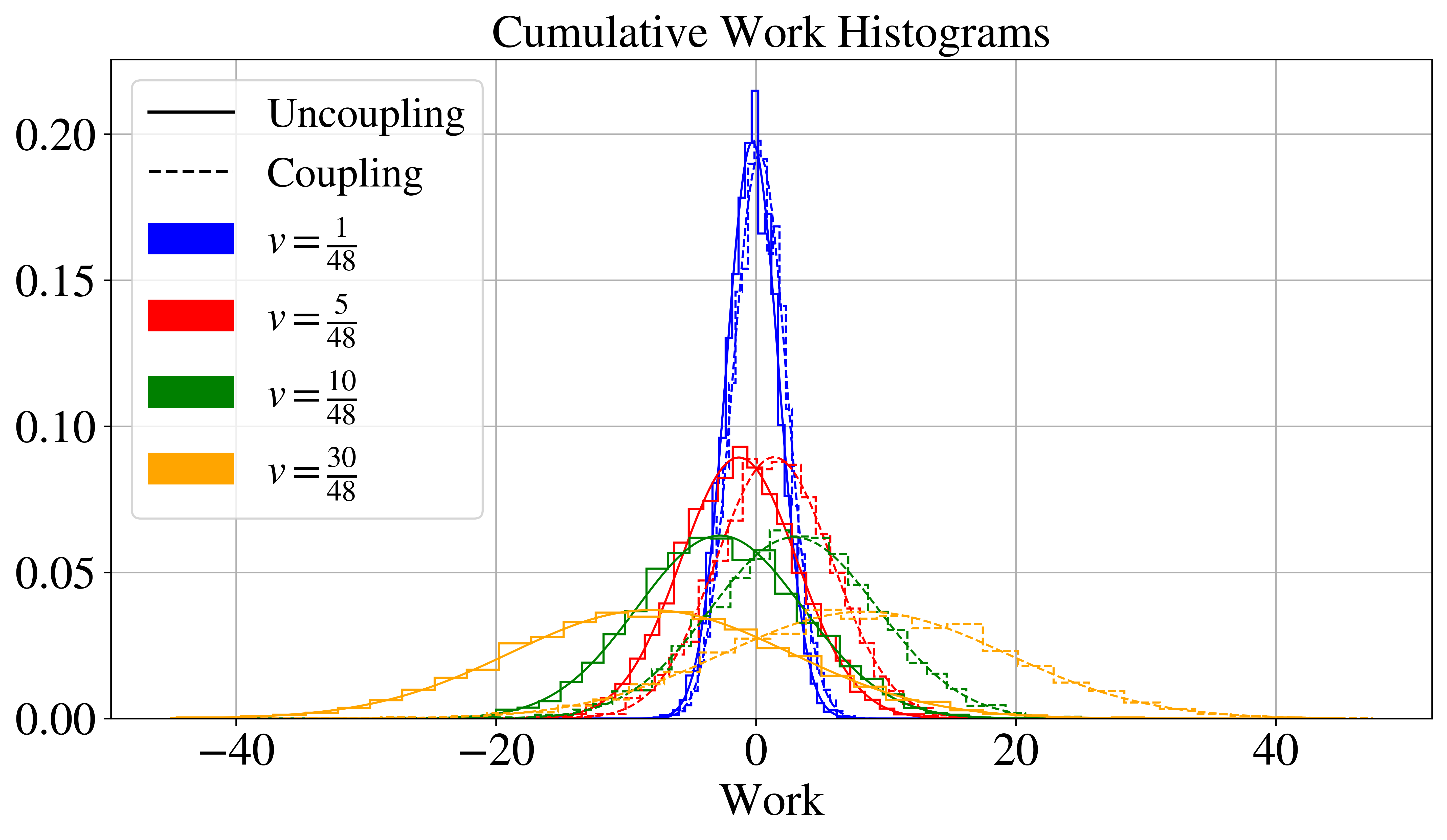}
    \caption{Histograms of work done in the coupling and uncoupling protocols in the second model for different velocities.}
    \label{fig:work_distribution_system2}
\end{figure}

\begin{figure}[h]
    \centering
    \includegraphics[width=0.42\textwidth]{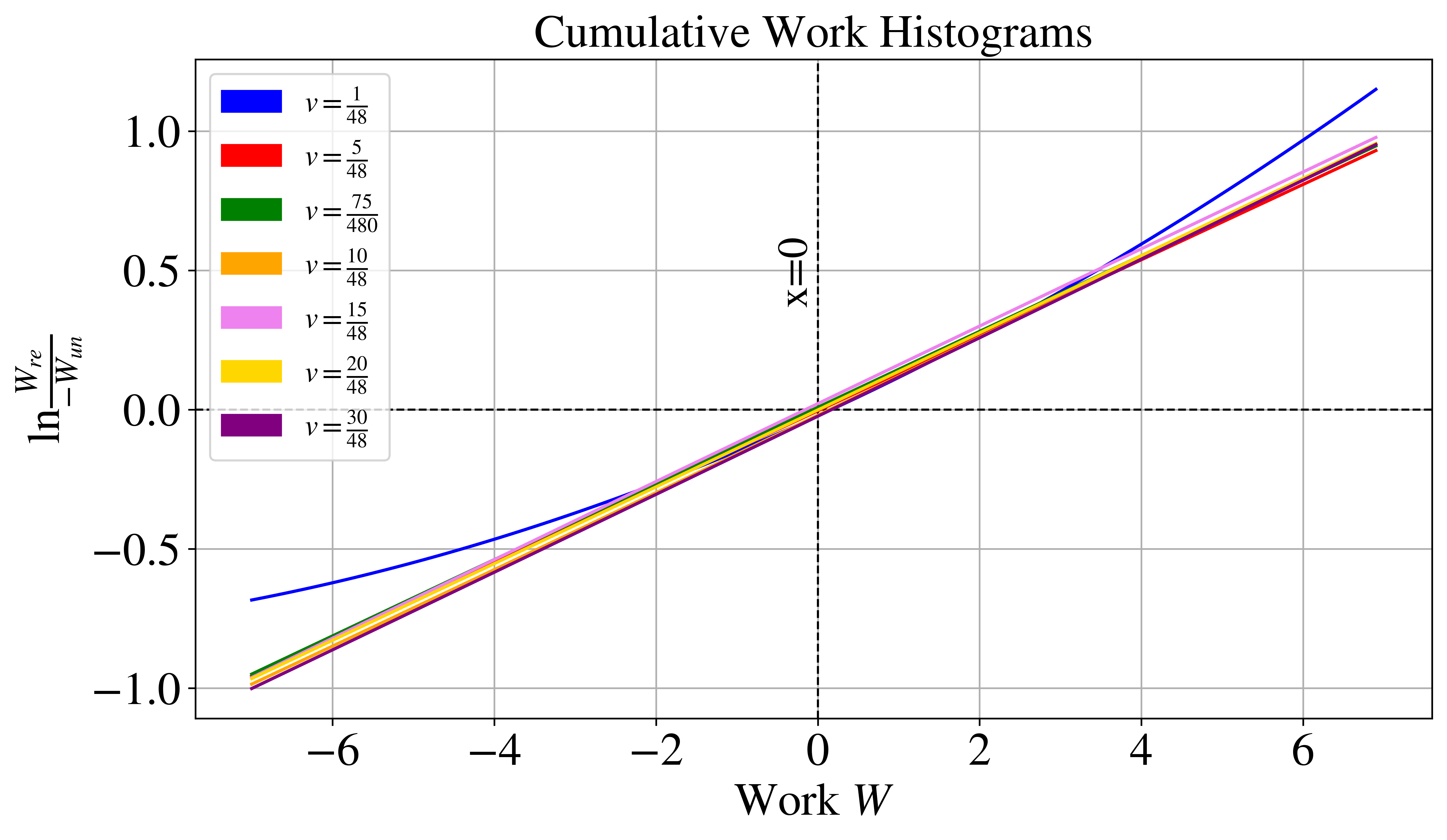}
    \caption{Crooks theorem for many velocities in the second model.}
    \label{fig:Crooks_system2}
\end{figure}

Finally, Fig. \ref{fig:work_distribution_system2} shows the work histograms obtained at several speeds. The intersecting point for forward and backwards histograms gives $\Delta F_\text{tot} \simeq 0.04 \pm 0.11$. In addition, Fig. \ref{fig:Crooks_system2} shows the linear relation between $\ln \frac{P_\text{for}}{P_\text{rev}}$ and $\Delta W$ that allows to use the CFT (Eq. \eqref{eq:crooks}) to find a free energy difference of $\Delta F_\text{tot} = 0.12\pm 0.40$. Both results agree with the theoretical null value for the free energy difference.

The results for both models validate CFT as an efficient method to estimate the free energy difference from nonequilibrium protocols, also in simulations.

\section{\label{conclusions}Conclusions}

The Crooks Fluctuation Theorem has proven to be a valuable tool for measuring free energy differences in experiments where a single molecule is pulled apart and pushed together. In this work, we introduce and simulate by using Brownian dynamics two mechanical models of coupled harmonic oscillators that reproduce the instantaneous force profiles typical of those single-molecule experiments. We verify that the Crooks relation holds, also for nonequilibrium processes, as expected. Even more, the force hysteresis curve for forward and backward protocols in one of the models (where a joining springs breaks and rejoins) is similar to the one reported by Collin, et al.~\cite{Bustamante_Crooks} when pulling a single ARN molecule. Thus, we evidence that such simplified models capture the essence of those experimental phenomena.

Future works may use these kind of mechanical models, not just to study similar experiments such as DNA torque detection~\cite{deufelNanofabricatedQuartzCylinders2007}, but also to reproduce the mechanical behavior of molecular motors such as ATP synthase~\cite{Sivak_F1-ATPase_optimal_control} or ARN polymerases~\cite{RNAPolymerasesMolecular2021}. Moreover, the simple construction of these mechanical systems renders them excellent tools to explain and discuss with students learning about stochastic thermodynamics and fluctuation theorems from a theoretical and computational standpoint. Additionally, computing other thermodynamic quantities such as heat and entropy could also be instructive. Finally, varying the proposed protocols could also allow students to compare different classical thermodynamic systems with their stochastic counterparts, e.g., to reproduce Brownian thermodynamic cycles such as Carnot or Stirling Brownian engines~\cite{martinezBrownianCarnotEngine2016, prieto-rodriguezMaximumPowerStirlinglikeHeat2025}.

This work has shown that simple mechanical models under Brownian dynamics are able to capture the essential behavior of complex single-molecule experiments. Such mechanical systems can also be used for teaching stochastic thermodynamics and fluctuation theorems, both from a theoretical and numerical point of view. That make those models a valuable tool for enlightening how fluctuation theorems work in such systems.

\clearpage

\bibliography{references}

\begin{thebibliography}{27}%
\makeatletter
\providecommand \@ifxundefined [1]{%
 \@ifx{#1\undefined}
}%
\providecommand \@ifnum [1]{%
 \ifnum #1\expandafter \@firstoftwo
 \else \expandafter \@secondoftwo
 \fi
}%
\providecommand \@ifx [1]{%
 \ifx #1\expandafter \@firstoftwo
 \else \expandafter \@secondoftwo
 \fi
}%
\providecommand \natexlab [1]{#1}%
\providecommand \enquote  [1]{``#1''}%
\providecommand \bibnamefont  [1]{#1}%
\providecommand \bibfnamefont [1]{#1}%
\providecommand \citenamefont [1]{#1}%
\providecommand \href@noop [0]{\@secondoftwo}%
\providecommand \href [0]{\begingroup \@sanitize@url \@href}%
\providecommand \@href[1]{\@@startlink{#1}\@@href}%
\providecommand \@@href[1]{\endgroup#1\@@endlink}%
\providecommand \@sanitize@url [0]{\catcode `\\12\catcode `\$12\catcode `\&12\catcode `\#12\catcode `\^12\catcode `\_12\catcode `\%12\relax}%
\providecommand \@@startlink[1]{}%
\providecommand \@@endlink[0]{}%
\providecommand \url  [0]{\begingroup\@sanitize@url \@url }%
\providecommand \@url [1]{\endgroup\@href {#1}{\urlprefix }}%
\providecommand \urlprefix  [0]{URL }%
\providecommand \Eprint [0]{\href }%
\providecommand \doibase [0]{https://doi.org/}%
\providecommand \selectlanguage [0]{\@gobble}%
\providecommand \bibinfo  [0]{\@secondoftwo}%
\providecommand \bibfield  [0]{\@secondoftwo}%
\providecommand \translation [1]{[#1]}%
\providecommand \BibitemOpen [0]{}%
\providecommand \bibitemStop [0]{}%
\providecommand \bibitemNoStop [0]{.\EOS\space}%
\providecommand \EOS [0]{\spacefactor3000\relax}%
\providecommand \BibitemShut  [1]{\csname bibitem#1\endcsname}%
\let\auto@bib@innerbib\@empty
%</preamble>
\bibitem [{\citenamefont {Sekimoto}(1998)}]{Sekimoto}%
  \BibitemOpen
  \bibfield  {author} {\bibinfo {author} {\bibfnamefont {K.}~\bibnamefont {Sekimoto}},\ }\bibfield  {title} {\bibinfo {title} {{Langevin Equation and Thermodynamics}},\ }\href {https://doi.org/10.1143/PTPS.130.17} {\bibfield  {journal} {\bibinfo  {journal} {Progress of Theoretical Physics Supplement}\ }\textbf {\bibinfo {volume} {130}},\ \bibinfo {pages} {17} (\bibinfo {year} {1998})}\BibitemShut {NoStop}%
\bibitem [{\citenamefont {Bhat}\ \emph {et~al.}(2022)\citenamefont {Bhat}, \citenamefont {Hauf}, \citenamefont {Plessy}, \citenamefont {Yokobayashi},\ and\ \citenamefont {Pigolotti}}]{bacterial_replisomes}%
  \BibitemOpen
  \bibfield  {author} {\bibinfo {author} {\bibfnamefont {D.}~\bibnamefont {Bhat}}, \bibinfo {author} {\bibfnamefont {S.}~\bibnamefont {Hauf}}, \bibinfo {author} {\bibfnamefont {C.}~\bibnamefont {Plessy}}, \bibinfo {author} {\bibfnamefont {Y.}~\bibnamefont {Yokobayashi}},\ and\ \bibinfo {author} {\bibfnamefont {S.}~\bibnamefont {Pigolotti}},\ }\bibfield  {title} {\bibinfo {title} {Speed variations of bacterial replisomes},\ }\href {https://doi.org/10.7554/eLife.75884} {\bibfield  {journal} {\bibinfo  {journal} {eLife}\ }\textbf {\bibinfo {volume} {11}},\ \bibinfo {pages} {e75884} (\bibinfo {year} {2022})}\BibitemShut {NoStop}%
\bibitem [{\citenamefont {Lu}\ \emph {et~al.}(2021)\citenamefont {Lu}, \citenamefont {Bhat}, \citenamefont {Stepanenko},\ and\ \citenamefont {Pigolotti}}]{CRISPR_Dynamics}%
  \BibitemOpen
  \bibfield  {author} {\bibinfo {author} {\bibfnamefont {Q.}~\bibnamefont {Lu}}, \bibinfo {author} {\bibfnamefont {D.}~\bibnamefont {Bhat}}, \bibinfo {author} {\bibfnamefont {D.}~\bibnamefont {Stepanenko}},\ and\ \bibinfo {author} {\bibfnamefont {S.}~\bibnamefont {Pigolotti}},\ }\bibfield  {title} {\bibinfo {title} {Search and localization dynamics of the crispr-cas9 system},\ }\href {https://doi.org/10.1103/PhysRevLett.127.208102} {\bibfield  {journal} {\bibinfo  {journal} {Phys. Rev. Lett.}\ }\textbf {\bibinfo {volume} {127}},\ \bibinfo {pages} {208102} (\bibinfo {year} {2021})}\BibitemShut {NoStop}%
\bibitem [{\citenamefont {Blaber}\ and\ \citenamefont {Sivak}(2020)}]{Sivak_protein_copy-number}%
  \BibitemOpen
  \bibfield  {author} {\bibinfo {author} {\bibfnamefont {S.}~\bibnamefont {Blaber}}\ and\ \bibinfo {author} {\bibfnamefont {D.~A.}\ \bibnamefont {Sivak}},\ }\bibfield  {title} {\bibinfo {title} {Optimal control of protein copy number},\ }\href {https://doi.org/10.1103/PhysRevE.101.022118} {\bibfield  {journal} {\bibinfo  {journal} {Phys. Rev. E}\ }\textbf {\bibinfo {volume} {101}},\ \bibinfo {pages} {022118} (\bibinfo {year} {2020})}\BibitemShut {NoStop}%
\bibitem [{\citenamefont {Gupta}\ \emph {et~al.}(2022)\citenamefont {Gupta}, \citenamefont {Large}, \citenamefont {Toyabe},\ and\ \citenamefont {Sivak}}]{Sivak_F1-ATPase_optimal_control}%
  \BibitemOpen
  \bibfield  {author} {\bibinfo {author} {\bibfnamefont {D.}~\bibnamefont {Gupta}}, \bibinfo {author} {\bibfnamefont {S.~J.}\ \bibnamefont {Large}}, \bibinfo {author} {\bibfnamefont {S.}~\bibnamefont {Toyabe}},\ and\ \bibinfo {author} {\bibfnamefont {D.~A.}\ \bibnamefont {Sivak}},\ }\bibfield  {title} {\bibinfo {title} {Optimal control of the f1-atpase molecular motor},\ }\href {https://doi.org/10.1021/acs.jpclett.2c03033} {\bibfield  {journal} {\bibinfo  {journal} {The Journal of Physical Chemistry Letters}\ }\textbf {\bibinfo {volume} {13}},\ \bibinfo {pages} {11844} (\bibinfo {year} {2022})}\BibitemShut {NoStop}%
\bibitem [{\citenamefont {Seifert}(2008)}]{Seifert_Principles}%
  \BibitemOpen
  \bibfield  {author} {\bibinfo {author} {\bibfnamefont {U.}~\bibnamefont {Seifert}},\ }\bibfield  {title} {\bibinfo {title} {Stochastic thermodynamics: principles and perspectives},\ }\href {https://doi.org/10.1140/epjb/e2008-00001-9} {\bibfield  {journal} {\bibinfo  {journal} {The European Physics Journal B}\ }\textbf {\bibinfo {volume} {64}},\ \bibinfo {pages} {423} (\bibinfo {year} {2008})}\BibitemShut {NoStop}%
\bibitem [{\citenamefont {Evans}\ and\ \citenamefont {Searles}(1994)}]{DetailedFT_Evans_Searles}%
  \BibitemOpen
  \bibfield  {author} {\bibinfo {author} {\bibfnamefont {D.~J.}\ \bibnamefont {Evans}}\ and\ \bibinfo {author} {\bibfnamefont {D.~J.}\ \bibnamefont {Searles}},\ }\bibfield  {title} {\bibinfo {title} {Equilibrium microstates which generate second law violating steady states},\ }\href {https://doi.org/10.1103/PhysRevE.50.1645} {\bibfield  {journal} {\bibinfo  {journal} {Phys. Rev. E}\ }\textbf {\bibinfo {volume} {50}},\ \bibinfo {pages} {1645} (\bibinfo {year} {1994})}\BibitemShut {NoStop}%
\bibitem [{\citenamefont {Jarzynski}(1997{\natexlab{a}})}]{Jarzynski1}%
  \BibitemOpen
  \bibfield  {author} {\bibinfo {author} {\bibfnamefont {C.}~\bibnamefont {Jarzynski}},\ }\bibfield  {title} {\bibinfo {title} {Nonequilibrium equality for free energy differences},\ }\href {https://doi.org/10.1103/PhysRevLett.78.2690} {\bibfield  {journal} {\bibinfo  {journal} {Phys. Rev. Lett.}\ }\textbf {\bibinfo {volume} {78}},\ \bibinfo {pages} {2690} (\bibinfo {year} {1997}{\natexlab{a}})}\BibitemShut {NoStop}%
\bibitem [{\citenamefont {Jarzynski}(1997{\natexlab{b}})}]{Jarzynski2}%
  \BibitemOpen
  \bibfield  {author} {\bibinfo {author} {\bibfnamefont {C.}~\bibnamefont {Jarzynski}},\ }\bibfield  {title} {\bibinfo {title} {Equilibrium free-energy differences from nonequilibrium measurements: A master-equation approach},\ }\href {https://doi.org/10.1103/PhysRevE.56.5018} {\bibfield  {journal} {\bibinfo  {journal} {Phys. Rev. E}\ }\textbf {\bibinfo {volume} {56}},\ \bibinfo {pages} {5018} (\bibinfo {year} {1997}{\natexlab{b}})}\BibitemShut {NoStop}%
\bibitem [{\citenamefont {Crooks}(1999)}]{Crooks}%
  \BibitemOpen
  \bibfield  {author} {\bibinfo {author} {\bibfnamefont {G.~E.}\ \bibnamefont {Crooks}},\ }\bibfield  {title} {\bibinfo {title} {Entropy production fluctuation theorem and the nonequilibrium work relation for free energy differences},\ }\href {https://doi.org/10.1103/PhysRevE.60.2721} {\bibfield  {journal} {\bibinfo  {journal} {Phys. Rev. E}\ }\textbf {\bibinfo {volume} {60}},\ \bibinfo {pages} {2721} (\bibinfo {year} {1999})}\BibitemShut {NoStop}%
\bibitem [{\citenamefont {Seifert}(2005)}]{seifert_IFT}%
  \BibitemOpen
  \bibfield  {author} {\bibinfo {author} {\bibfnamefont {U.}~\bibnamefont {Seifert}},\ }\bibfield  {title} {\bibinfo {title} {Entropy production along a stochastic trajectory and an integral fluctuation theorem},\ }\href {https://doi.org/10.1103/PhysRevLett.95.040602} {\bibfield  {journal} {\bibinfo  {journal} {Phys. Rev. Lett.}\ }\textbf {\bibinfo {volume} {95}},\ \bibinfo {pages} {040602} (\bibinfo {year} {2005})}\BibitemShut {NoStop}%
\bibitem [{\citenamefont {Hummer}\ and\ \citenamefont {Szabo}(2001)}]{Hummer_Szabo}%
  \BibitemOpen
  \bibfield  {author} {\bibinfo {author} {\bibfnamefont {G.}~\bibnamefont {Hummer}}\ and\ \bibinfo {author} {\bibfnamefont {A.}~\bibnamefont {Szabo}},\ }\bibfield  {title} {\bibinfo {title} {Free energy reconstruction from nonequilibrium single-molecule pulling experiments},\ }\href {https://doi.org/10.1073/pnas.071034098} {\bibfield  {journal} {\bibinfo  {journal} {Proceedings of the National Academy of Sciences}\ }\textbf {\bibinfo {volume} {98}},\ \bibinfo {pages} {3658–3661} (\bibinfo {year} {2001})}\BibitemShut {NoStop}%
\bibitem [{\citenamefont {Liphardt}\ \emph {et~al.}(2002)\citenamefont {Liphardt}, \citenamefont {Dumont}, \citenamefont {Smith}, \citenamefont {Tinoco},\ and\ \citenamefont {Bustamante}}]{bustamante_jarzynski}%
  \BibitemOpen
  \bibfield  {author} {\bibinfo {author} {\bibfnamefont {J.}~\bibnamefont {Liphardt}}, \bibinfo {author} {\bibfnamefont {S.}~\bibnamefont {Dumont}}, \bibinfo {author} {\bibfnamefont {S.~B.}\ \bibnamefont {Smith}}, \bibinfo {author} {\bibfnamefont {I.}~\bibnamefont {Tinoco}},\ and\ \bibinfo {author} {\bibfnamefont {C.}~\bibnamefont {Bustamante}},\ }\bibfield  {title} {\bibinfo {title} {Equilibrium information from nonequilibrium measurements in an experimental test of jarzynski's equality},\ }\href {https://doi.org/10.1126/science.1071152} {\bibfield  {journal} {\bibinfo  {journal} {Science}\ }\textbf {\bibinfo {volume} {296}},\ \bibinfo {pages} {1832} (\bibinfo {year} {2002})}\BibitemShut {NoStop}%
\bibitem [{\citenamefont {Collin}\ \emph {et~al.}(2005)\citenamefont {Collin}, \citenamefont {Ritort}, \citenamefont {Jarzynski}, \citenamefont {Smith}, \citenamefont {Tinoco},\ and\ \citenamefont {Bustamante}}]{Bustamante_Crooks}%
  \BibitemOpen
  \bibfield  {author} {\bibinfo {author} {\bibfnamefont {D.}~\bibnamefont {Collin}}, \bibinfo {author} {\bibfnamefont {F.}~\bibnamefont {Ritort}}, \bibinfo {author} {\bibfnamefont {C.}~\bibnamefont {Jarzynski}}, \bibinfo {author} {\bibfnamefont {S.~B.}\ \bibnamefont {Smith}}, \bibinfo {author} {\bibfnamefont {I.}~\bibnamefont {Tinoco}},\ and\ \bibinfo {author} {\bibfnamefont {C.}~\bibnamefont {Bustamante}},\ }\bibfield  {title} {\bibinfo {title} {Verification of the {{Crooks}} fluctuation theorem and recovery of {{RNA}} folding free energies},\ }\href {https://doi.org/10.1038/nature04061} {\bibfield  {journal} {\bibinfo  {journal} {Nature}\ }\textbf {\bibinfo {volume} {437}},\ \bibinfo {pages} {231} (\bibinfo {year} {2005})}\BibitemShut {NoStop}%
\bibitem [{\citenamefont {Severino}\ \emph {et~al.}(2019)\citenamefont {Severino}, \citenamefont {Monge}, \citenamefont {Rissone},\ and\ \citenamefont {Ritort}}]{severino_single-molecule_pulling_experiments}%
  \BibitemOpen
  \bibfield  {author} {\bibinfo {author} {\bibfnamefont {A.}~\bibnamefont {Severino}}, \bibinfo {author} {\bibfnamefont {A.~M.}\ \bibnamefont {Monge}}, \bibinfo {author} {\bibfnamefont {P.}~\bibnamefont {Rissone}},\ and\ \bibinfo {author} {\bibfnamefont {F.}~\bibnamefont {Ritort}},\ }\bibfield  {title} {\bibinfo {title} {Efficient methods for determining folding free energies in single-molecule pulling experiments},\ }\href {https://doi.org/10.1088/1742-5468/ab4e91} {\bibfield  {journal} {\bibinfo  {journal} {Journal of Statistical Mechanics: Theory and Experiment}\ }\textbf {\bibinfo {volume} {2019}},\ \bibinfo {pages} {124001} (\bibinfo {year} {2019})}\BibitemShut {NoStop}%
\bibitem [{\citenamefont {Calzetta}(2009)}]{kinesin_crooks_FT}%
  \BibitemOpen
  \bibfield  {author} {\bibinfo {author} {\bibfnamefont {E.~A.}\ \bibnamefont {Calzetta}},\ }\bibfield  {title} {\bibinfo {title} {Kinesin and the {C}rooks fluctuation theorem},\ }\href {https://doi.org/10.1140/epjb/e2009-00113-8} {\bibfield  {journal} {\bibinfo  {journal} {The European Physical Journal B}\ }\textbf {\bibinfo {volume} {68}},\ \bibinfo {pages} {601} (\bibinfo {year} {2009})}\BibitemShut {NoStop}%
\bibitem [{\citenamefont {Hayashi}\ \emph {et~al.}(2010)\citenamefont {Hayashi}, \citenamefont {Ueno}, \citenamefont {Iino},\ and\ \citenamefont {Noji}}]{Fluctuation_Theorem_F1-ATPase}%
  \BibitemOpen
  \bibfield  {author} {\bibinfo {author} {\bibfnamefont {K.}~\bibnamefont {Hayashi}}, \bibinfo {author} {\bibfnamefont {H.}~\bibnamefont {Ueno}}, \bibinfo {author} {\bibfnamefont {R.}~\bibnamefont {Iino}},\ and\ \bibinfo {author} {\bibfnamefont {H.}~\bibnamefont {Noji}},\ }\bibfield  {title} {\bibinfo {title} {Fluctuation theorem applied to ${\mathbf{f}}_{1}$-atpase},\ }\href {https://doi.org/10.1103/PhysRevLett.104.218103} {\bibfield  {journal} {\bibinfo  {journal} {Phys. Rev. Lett.}\ }\textbf {\bibinfo {volume} {104}},\ \bibinfo {pages} {218103} (\bibinfo {year} {2010})}\BibitemShut {NoStop}%
\bibitem [{\citenamefont {Abraham}\ \emph {et~al.}(2023)\citenamefont {Abraham}, \citenamefont {Alekseenko}, \citenamefont {Bergh}, \citenamefont {Blau}, \citenamefont {Briand}, \citenamefont {Doijade}, \citenamefont {Fleischmann}, \citenamefont {Gapsys}, \citenamefont {Garg}, \citenamefont {Gorelov}, \citenamefont {Gouaillardet}, \citenamefont {Gray}, \citenamefont {Eric~Irrgang}, \citenamefont {Jalalypour}, \citenamefont {Jordan}, \citenamefont {Junghans}, \citenamefont {Kanduri}, \citenamefont {Keller}, \citenamefont {Kutzner}, \citenamefont {Lemkul}, \citenamefont {Lundborg}, \citenamefont {Merz}, \citenamefont {Mileti{\'c}}, \citenamefont {Morozov}, \citenamefont {P{\'a}ll}, \citenamefont {Schulz}, \citenamefont {Shirts}, \citenamefont {Shvetsov}, \citenamefont {Soproni}, \citenamefont {van~der Spoel}, \citenamefont {Turner}, \citenamefont {Uphoff}, \citenamefont {Villa}, \citenamefont {Wingberm{\"u}hle}, \citenamefont {Zhmurov}, \citenamefont {Bauer}, \citenamefont {Hess},\ and\ \citenamefont {Lindahl}}]{gromacs}%
  \BibitemOpen
  \bibfield  {author} {\bibinfo {author} {\bibfnamefont {M.}~\bibnamefont {Abraham}}, \bibinfo {author} {\bibfnamefont {A.}~\bibnamefont {Alekseenko}}, \bibinfo {author} {\bibfnamefont {C.}~\bibnamefont {Bergh}}, \bibinfo {author} {\bibfnamefont {C.}~\bibnamefont {Blau}}, \bibinfo {author} {\bibfnamefont {E.}~\bibnamefont {Briand}}, \bibinfo {author} {\bibfnamefont {M.}~\bibnamefont {Doijade}}, \bibinfo {author} {\bibfnamefont {S.}~\bibnamefont {Fleischmann}}, \bibinfo {author} {\bibfnamefont {V.}~\bibnamefont {Gapsys}}, \bibinfo {author} {\bibfnamefont {G.}~\bibnamefont {Garg}}, \bibinfo {author} {\bibfnamefont {S.}~\bibnamefont {Gorelov}}, \bibinfo {author} {\bibfnamefont {G.}~\bibnamefont {Gouaillardet}}, \bibinfo {author} {\bibfnamefont {A.}~\bibnamefont {Gray}}, \bibinfo {author} {\bibfnamefont {M.}~\bibnamefont {Eric~Irrgang}}, \bibinfo {author} {\bibfnamefont {F.}~\bibnamefont {Jalalypour}}, \bibinfo {author} {\bibfnamefont {J.}~\bibnamefont {Jordan}}, \bibinfo {author} {\bibfnamefont {C.}~\bibnamefont {Junghans}}, \bibinfo {author} {\bibfnamefont {P.}~\bibnamefont {Kanduri}}, \bibinfo {author} {\bibfnamefont {S.}~\bibnamefont {Keller}}, \bibinfo {author} {\bibfnamefont {C.}~\bibnamefont {Kutzner}}, \bibinfo {author} {\bibfnamefont {J.~A.}\ \bibnamefont {Lemkul}}, \bibinfo {author} {\bibfnamefont {M.}~\bibnamefont {Lundborg}}, \bibinfo {author} {\bibfnamefont {P.}~\bibnamefont {Merz}}, \bibinfo {author} {\bibfnamefont {V.}~\bibnamefont {Mileti{\'c}}}, \bibinfo {author} {\bibfnamefont {D.}~\bibnamefont {Morozov}}, \bibinfo {author} {\bibfnamefont {S.}~\bibnamefont {P{\'a}ll}}, \bibinfo {author} {\bibfnamefont {R.}~\bibnamefont {Schulz}}, \bibinfo {author} {\bibfnamefont {M.}~\bibnamefont {Shirts}}, \bibinfo {author} {\bibfnamefont {A.}~\bibnamefont {Shvetsov}}, \bibinfo {author} {\bibfnamefont {B.}~\bibnamefont {Soproni}}, \bibinfo {author} {\bibfnamefont {D.}~\bibnamefont {van~der Spoel}}, \bibinfo {author} {\bibfnamefont {P.}~\bibnamefont {Turner}}, \bibinfo {author} {\bibfnamefont {C.}~\bibnamefont {Uphoff}}, \bibinfo {author} {\bibfnamefont {A.}~\bibnamefont {Villa}}, \bibinfo {author} {\bibfnamefont {S.}~\bibnamefont {Wingberm{\"u}hle}}, \bibinfo {author} {\bibfnamefont {A.}~\bibnamefont {Zhmurov}}, \bibinfo {author} {\bibfnamefont {P.}~\bibnamefont {Bauer}}, \bibinfo {author} {\bibfnamefont {B.}~\bibnamefont {Hess}},\ and\ \bibinfo {author} {\bibfnamefont {E.}~\bibnamefont {Lindahl}},\ }\href@noop {} {\bibinfo {title} {{GROMACS} 2023.3 manual}} (\bibinfo {year} {2023})\BibitemShut {NoStop}%
\bibitem [{\citenamefont {Brooks}\ \emph {et~al.}(2009)\citenamefont {Brooks}, \citenamefont {Brooks}, \citenamefont {Mackerell}, \citenamefont {Nilsson}, \citenamefont {Petrella}, \citenamefont {Roux}, \citenamefont {Won}, \citenamefont {Archontis}, \citenamefont {Bartels}, \citenamefont {Boresch}, \citenamefont {Caflisch}, \citenamefont {Caves}, \citenamefont {Cui}, \citenamefont {Dinner}, \citenamefont {Feig}, \citenamefont {Fischer}, \citenamefont {Gao}, \citenamefont {Hodoscek}, \citenamefont {Im}, \citenamefont {Kuczera}, \citenamefont {Lazaridis}, \citenamefont {Ma}, \citenamefont {Ovchinnikov}, \citenamefont {Paci}, \citenamefont {Pastor}, \citenamefont {Post}, \citenamefont {Pu}, \citenamefont {Schaefer}, \citenamefont {Tidor}, \citenamefont {Venable}, \citenamefont {Woodcock}, \citenamefont {Wu}, \citenamefont {Yang}, \citenamefont {York},\ and\ \citenamefont {Karplus}}]{charmm}%
  \BibitemOpen
  \bibfield  {author} {\bibinfo {author} {\bibfnamefont {B.~R.}\ \bibnamefont {Brooks}}, \bibinfo {author} {\bibfnamefont {C.~L.}\ \bibnamefont {Brooks}}, \bibinfo {author} {\bibfnamefont {A.~D.}\ \bibnamefont {Mackerell}}, \bibinfo {author} {\bibfnamefont {L.}~\bibnamefont {Nilsson}}, \bibinfo {author} {\bibfnamefont {R.~J.}\ \bibnamefont {Petrella}}, \bibinfo {author} {\bibfnamefont {B.}~\bibnamefont {Roux}}, \bibinfo {author} {\bibfnamefont {Y.}~\bibnamefont {Won}}, \bibinfo {author} {\bibfnamefont {G.}~\bibnamefont {Archontis}}, \bibinfo {author} {\bibfnamefont {C.}~\bibnamefont {Bartels}}, \bibinfo {author} {\bibfnamefont {S.}~\bibnamefont {Boresch}}, \bibinfo {author} {\bibfnamefont {A.}~\bibnamefont {Caflisch}}, \bibinfo {author} {\bibfnamefont {L.}~\bibnamefont {Caves}}, \bibinfo {author} {\bibfnamefont {Q.}~\bibnamefont {Cui}}, \bibinfo {author} {\bibfnamefont {A.~R.}\ \bibnamefont {Dinner}}, \bibinfo {author} {\bibfnamefont {M.}~\bibnamefont {Feig}}, \bibinfo {author} {\bibfnamefont {S.}~\bibnamefont {Fischer}}, \bibinfo {author} {\bibfnamefont {J.}~\bibnamefont {Gao}}, \bibinfo {author} {\bibfnamefont {M.}~\bibnamefont {Hodoscek}}, \bibinfo {author} {\bibfnamefont {W.}~\bibnamefont {Im}}, \bibinfo {author} {\bibfnamefont {K.}~\bibnamefont {Kuczera}}, \bibinfo {author} {\bibfnamefont {T.}~\bibnamefont {Lazaridis}}, \bibinfo {author} {\bibfnamefont {J.}~\bibnamefont {Ma}}, \bibinfo {author} {\bibfnamefont {V.}~\bibnamefont {Ovchinnikov}}, \bibinfo {author} {\bibfnamefont {E.}~\bibnamefont {Paci}}, \bibinfo {author} {\bibfnamefont {R.~W.}\ \bibnamefont {Pastor}}, \bibinfo {author} {\bibfnamefont {C.~B.}\ \bibnamefont {Post}}, \bibinfo {author} {\bibfnamefont {J.~Z.}\ \bibnamefont {Pu}}, \bibinfo {author} {\bibfnamefont {M.}~\bibnamefont {Schaefer}}, \bibinfo {author} {\bibfnamefont {B.}~\bibnamefont {Tidor}}, \bibinfo {author} {\bibfnamefont {R.~M.}\ \bibnamefont {Venable}}, \bibinfo {author} {\bibfnamefont {H.~L.}\ \bibnamefont {Woodcock}}, \bibinfo {author} {\bibfnamefont {X.}~\bibnamefont {Wu}}, \bibinfo {author} {\bibfnamefont {W.}~\bibnamefont {Yang}}, \bibinfo {author} {\bibfnamefont {D.~M.}\ \bibnamefont {York}},\ and\ \bibinfo {author} {\bibfnamefont {M.}~\bibnamefont {Karplus}},\ }\bibfield  {title} {\bibinfo {title} {{{CHARMM}}: {{The}} biomolecular simulation program},\ }\href {https://doi.org/10.1002/jcc.21287} {\bibfield  {journal} {\bibinfo  {journal} {Journal of Computational Chemistry}\ }\textbf {\bibinfo {volume} {30}},\ \bibinfo {pages} {1545} (\bibinfo {year} {2009})}\BibitemShut {NoStop}%
\bibitem [{\citenamefont {Kramers}(1940)}]{Kramers}%
  \BibitemOpen
  \bibfield  {author} {\bibinfo {author} {\bibfnamefont {H.}~\bibnamefont {Kramers}},\ }\bibfield  {title} {\bibinfo {title} {Brownian motion in a field of force and the diffusion model of chemical reactions},\ }\href {https://doi.org/https://doi.org/10.1016/S0031-8914(40)90098-2} {\bibfield  {journal} {\bibinfo  {journal} {Physica}\ }\textbf {\bibinfo {volume} {7}},\ \bibinfo {pages} {284} (\bibinfo {year} {1940})}\BibitemShut {NoStop}%
\bibitem [{\citenamefont {Moyal}(1949)}]{Moyal}%
  \BibitemOpen
  \bibfield  {author} {\bibinfo {author} {\bibfnamefont {J.~E.}\ \bibnamefont {Moyal}},\ }\bibfield  {title} {\bibinfo {title} {Stochastic processes and statistical physics},\ }\href {http://www.jstor.org/stable/2984076} {\bibfield  {journal} {\bibinfo  {journal} {Journal of the Royal Statistical Society. Series B (Methodological)}\ }\textbf {\bibinfo {volume} {11}},\ \bibinfo {pages} {150} (\bibinfo {year} {1949})}\BibitemShut {NoStop}%
\bibitem [{\citenamefont {Uhlenbeck}\ and\ \citenamefont {Ornstein}(1930)}]{Ornstein_Uhlenbeck}%
  \BibitemOpen
  \bibfield  {author} {\bibinfo {author} {\bibfnamefont {G.~E.}\ \bibnamefont {Uhlenbeck}}\ and\ \bibinfo {author} {\bibfnamefont {L.~S.}\ \bibnamefont {Ornstein}},\ }\bibfield  {title} {\bibinfo {title} {On the theory of the brownian motion},\ }\href {https://doi.org/10.1103/PhysRev.36.823} {\bibfield  {journal} {\bibinfo  {journal} {Phys. Rev.}\ }\textbf {\bibinfo {volume} {36}},\ \bibinfo {pages} {823} (\bibinfo {year} {1930})}\BibitemShut {NoStop}%
\bibitem [{\citenamefont {Gunsteren}\ and\ \citenamefont {Berendsen}(1988)}]{Gunsteren_Berendsen1}%
  \BibitemOpen
  \bibfield  {author} {\bibinfo {author} {\bibfnamefont {W.~F.~V.}\ \bibnamefont {Gunsteren}}\ and\ \bibinfo {author} {\bibfnamefont {H.~J.~C.}\ \bibnamefont {Berendsen}},\ }\bibfield  {title} {\bibinfo {title} {A leap-frog algorithm for stochastic dynamics},\ }\href {https://doi.org/10.1080/08927028808080941} {\bibfield  {journal} {\bibinfo  {journal} {Molecular Simulation}\ }\textbf {\bibinfo {volume} {1}},\ \bibinfo {pages} {173} (\bibinfo {year} {1988})}\BibitemShut {NoStop}%
\bibitem [{\citenamefont {Deufel}\ \emph {et~al.}(2007)\citenamefont {Deufel}, \citenamefont {Forth}, \citenamefont {Simmons}, \citenamefont {Dejgosha},\ and\ \citenamefont {Wang}}]{deufelNanofabricatedQuartzCylinders2007}%
  \BibitemOpen
  \bibfield  {author} {\bibinfo {author} {\bibfnamefont {C.}~\bibnamefont {Deufel}}, \bibinfo {author} {\bibfnamefont {S.}~\bibnamefont {Forth}}, \bibinfo {author} {\bibfnamefont {C.~R.}\ \bibnamefont {Simmons}}, \bibinfo {author} {\bibfnamefont {S.}~\bibnamefont {Dejgosha}},\ and\ \bibinfo {author} {\bibfnamefont {M.~D.}\ \bibnamefont {Wang}},\ }\bibfield  {title} {\bibinfo {title} {Nanofabricated quartz cylinders for angular trapping: {{DNA}} supercoiling torque detection},\ }\href {https://doi.org/10.1038/nmeth1013} {\bibfield  {journal} {\bibinfo  {journal} {Nature Methods}\ }\textbf {\bibinfo {volume} {4}},\ \bibinfo {pages} {223} (\bibinfo {year} {2007})}\BibitemShut {NoStop}%
\bibitem [{RNA(2021)}]{RNAPolymerasesMolecular2021}%
  \BibitemOpen
  \href {https://doi.org/10.1039/9781839160561} {\emph {\bibinfo {title} {{{RNA Polymerases}} as {{Molecular Motors}}: {{On}} the {{Road}}}}}\ (\bibinfo  {publisher} {The Royal Society of Chemistry},\ \bibinfo {year} {2021})\BibitemShut {NoStop}%
\bibitem [{\citenamefont {Mart{\'i}nez}\ \emph {et~al.}(2016)\citenamefont {Mart{\'i}nez}, \citenamefont {Rold{\'a}n}, \citenamefont {Dinis}, \citenamefont {Petrov}, \citenamefont {Parrondo},\ and\ \citenamefont {Rica}}]{martinezBrownianCarnotEngine2016}%
  \BibitemOpen
  \bibfield  {author} {\bibinfo {author} {\bibfnamefont {I.~A.}\ \bibnamefont {Mart{\'i}nez}}, \bibinfo {author} {\bibfnamefont {{\'E}.}~\bibnamefont {Rold{\'a}n}}, \bibinfo {author} {\bibfnamefont {L.}~\bibnamefont {Dinis}}, \bibinfo {author} {\bibfnamefont {D.}~\bibnamefont {Petrov}}, \bibinfo {author} {\bibfnamefont {J.~M.~R.}\ \bibnamefont {Parrondo}},\ and\ \bibinfo {author} {\bibfnamefont {R.~A.}\ \bibnamefont {Rica}},\ }\bibfield  {title} {\bibinfo {title} {Brownian {{Carnot}} engine},\ }\href {https://doi.org/10.1038/nphys3518} {\bibfield  {journal} {\bibinfo  {journal} {Nature Physics}\ }\textbf {\bibinfo {volume} {12}},\ \bibinfo {pages} {67} (\bibinfo {year} {2016})}\BibitemShut {NoStop}%
\bibitem [{\citenamefont {{Prieto-Rodr{\'i}guez}}\ \emph {et~al.}(2025)\citenamefont {{Prieto-Rodr{\'i}guez}}, \citenamefont {Prados},\ and\ \citenamefont {Plata}}]{prieto-rodriguezMaximumPowerStirlinglikeHeat2025}%
  \BibitemOpen
  \bibfield  {author} {\bibinfo {author} {\bibfnamefont {I.}~\bibnamefont {{Prieto-Rodr{\'i}guez}}}, \bibinfo {author} {\bibfnamefont {A.}~\bibnamefont {Prados}},\ and\ \bibinfo {author} {\bibfnamefont {C.~A.}\ \bibnamefont {Plata}},\ }\bibfield  {title} {\bibinfo {title} {Maximum-{{Power Stirling-like Heat Engine}} with a {{Harmonically Confined Brownian Particle}}},\ }\href {https://doi.org/10.3390/e27010072} {\bibfield  {journal} {\bibinfo  {journal} {Entropy}\ }\textbf {\bibinfo {volume} {27}},\ \bibinfo {pages} {72} (\bibinfo {year} {2025})}\BibitemShut {NoStop}%
\end{thebibliography}%

\end{document}